\RequirePackage{lineno}
\documentclass[aps, prl,
floatfix,twocolumn,superscriptaddress,nofootinbib
]{revtex4-2}
\pdfoutput=1
\usepackage{t1enc}
\usepackage[utf8]{inputenc}
\usepackage[T1]{fontenc}
\usepackage{lmodern}
\usepackage{graphicx}
\usepackage{epsfig}
\usepackage{amssymb}
\usepackage{amsmath}
\usepackage{dcolumn}
\usepackage{bm}
\usepackage{color}
\usepackage{float}
\usepackage{hyperref}
\usepackage{natbib}
\usepackage{footnote}
\usepackage{ulem}
\usepackage{hhline}
\begin{document}

\title{About two-dimensional fits for the analysis of the scattering rates and renormalization functions from angle-resolved photoelectron spectroscopy data}

\author{R.\,Kurleto}
\affiliation {Leibniz Institute for Solid State and Materials Research  Dresden, Helmholtzstr. 20, D-01069 Dresden, Germany}
\affiliation {M. Smoluchowski Institute of Physics, Jagiellonian University, {\L}ojasiewicza 11, 30-348, Krak{\'o}w, Poland }
\affiliation{Department of Physics, University of Colorado, Boulder, CO, 80309, USA}

\author{J.\,Fink}
\affiliation{Leibniz Institute for Solid State and Materials Research  Dresden, Helmholtzstr. 20, D-01069 Dresden, Germany}
\affiliation {Institut f\"ur Festk\"orperphysik,  Technische Universität Dresden, D-01062 Dresden, Germany}

\begin{abstract}\
A new method for the analysis of the scattering rates from angle-resolved photoelectron spectroscopy (ARPES) is presented and described in details.  It takes into account experimental instrumental resolution and finite temperature effects. More accurate results are obtained in comparison with a standard, commonly used method. The application of the method is demonstrated for several examples commonly encountered in new quantum materials. Its usefulness is especially apparent in investigations of systems with strongly correlated electrons.
\end{abstract}


\maketitle

\section{Introduction}

The understanding of electronic correlation effects is one of the great  challenges of solid state physics. The reason is that many macroscopic properties such as electronic transport properties, thermal properties, magnetism, or superconductivity are related to these correlation effects.
\par
Experimentally the correlation effects can be studied by measuring for example the resistivity,  from which one can derive  the scattering rates, because they are at low temperatures mainly related to electron-electron interaction. The mass enhancement due to a polarization of the electron liquid can be derived for example from specific heat measurements at low temperatures.
This works well for simple metals which have very often just one band close to the chemical potential with an isotropic Fermi velocity. Most metallic materials, however, have a much more complicated electronic structure. There may be several bands close to the Fermi level and the mass renormalization of a band may be momentum dependent. In most cases, the above mentioned methods only provide a momentum averaged information. To obtain a microscopic understanding of the electronic structure of these more complicated systems one needs momentum dependent information. Angle-dependent photoemission spectroscopy (ARPES) is  a unique method which provides this momentum dependent information. Therefore it has developed to \textit{the} method to study the electronic structure of solids~\cite{sobota_review}.
\par
 ARPES measures the photocurrent, which is related to the spectral function $A(\omega,\mathbf{k},T)$ multiplied by a matrix element and by the temperature dependent Fermi function~\cite{Damascelli2003}. $\omega, \mathbf{k},$ and  $T$ is the binding energy relative to the Fermi level, the momentum, and the temperature, respectively. This product is then convoluted with the energy and momentum resolution. 
\par
The spectral function corresponds to the imaginary part of the Green's function which describes the propagation of a photo-hole~\cite{Mahan2000}. As a function of the renormalized band dispersion $\varepsilon^{\star}$ and as the scattering rate $\Gamma$ it can be written 
\begin{eqnarray}
A(\omega,\mathbf{k},T)=\frac{1}{\pi}Z\frac{\frac{\Gamma(\omega,\mathbf{k},T)}{2}}
{[\omega -\varepsilon^{\star}({\mathbf{k}})]^2+[\frac{\Gamma(\omega,\mathbf{k},T)}{2}]^2}.
\label{26_1}
\end{eqnarray}
Both, $\varepsilon^{\star}$ and $\Gamma$ can be related to the complex self-energy $\Sigma = \Re\Sigma + i \Im\Sigma$ ($\Re\Sigma$ and $\Im\Sigma$ are real and imaginary parts of the self-energy, respectively):
\begin{eqnarray}
  \varepsilon^{\star}({\mathbf{k})}=\varepsilon({\mathbf{k}})- \Re\Sigma(\omega,\mathbf{k},T)
\end{eqnarray}
where $\epsilon({\mathbf{k}})$ is the bare particle dispersion.
\begin{eqnarray}
 \Gamma(\omega,\mathbf{k},T)=-2Z(\omega,\mathbf{k},T) \Im\Sigma(\omega,\mathbf{k},T)
\end{eqnarray}
where Z  is the renormalization function directly related to the mass enhancement:
\begin{eqnarray}
Z=\frac{m}{m^{\star}}=\left(1- \frac{\partial\Re\Sigma}{\partial\omega}\right)^{-1}
\end{eqnarray}
This is valid for a one-band system for small self-energies close to the Fermi level, i.e., for a Fermi liquid when quasiparticles exist. Both, the imaginary part ($\Im\Sigma$) and the real part ($\Re\Sigma$) of $\Sigma$ are connected by the Kramers-Kronig transformation. However, one should have in mind that using the Kramers-Kronig transformation in case of a multi-band system is rather complicated due to interband interactions~\cite{ortenzi}.

In recent decades, the ARPES method has become a powerful tool in studies of solids. This is because, it delivers information about an electronic structure in a very direct way in the momentum space. Specific methods for analysis and interpretation of ARPES data have been also developed. The ARPES spectrum, i.e., a cut of the spectral weight along high symmetry direction in the momentum space collected at the given temperature $T$ can be analyzed in several ways. 

The most intuitive approach is to analyze sections corresponding to a fixed momentum $k$, i.e., energy distribution curves (EDC).  EDC profiles are Lorentzian peaks if the scattering rate $\Gamma$ is constant in energy. Otherwise, one needs to know the explicit form of the scattering rate $\Gamma(E)$ as a function of energy. The EDC analysis is especially useful in a case of weakly dispersive features in the electronic structure (quasiparticle bands in heavy fermion systems). However, one has to keep in mind, that EDC profiles are affected by the energy-dependent background due to additional scattering processes in photoemission.

The other approach is to analyze sections corresponding to a constant energy, i.e., momentum distribution curves (MDC). Assuming that the dispersion can be linearized, MDC profiles can be approximated by a Lorentzian function. Performing Lorentzian fits to MDCs, corresponding to subsequent energies $\omega$, one obtains the dispersion relation ($k$, $\omega=\varepsilon(k)$) from the maximum of the Lorentzian peaks. The scattering rates can be determined multiplying obtained Lorentzian full width at half maximum (FWHM) by the group velocity $v=\frac{d\varepsilon^{\star}}{dk}$ at a given energy. This method is free from detrimental influence of an energy dependent background and can be applied if the explicit form of $\Gamma(E)$ function is unknown. However, this method does not work properly if $\frac{d\varepsilon}{dk}\approx 0 $ (e.g. close to the top/bottom of the hole/electron pocket). If the dispersion consists of several branches then the spectrum is  fitted with  the linear combination of Lorentzian peaks. This is approximation, which works only if there is no big overlap between both branches. The evaluation of the scattering rates by the MDC method has been first applied to metals such as Be~\cite{LaShell2000}, Mo~\cite{Valla1999}, Ni~\cite{ni_scattering_rates}, Pd~\cite{pallad}, Al~\cite{al_kink}, Cu, and Pb~\cite{reinert_Pb_Cu}. The results of APRES lineshape analysis for noble metals are also reviewed in Ref.~\cite{MATZDORF}. Later also ARPES data from high $T_c$ superconductors have been evaluated using this method~\cite{Valla1999b,Kordyuk2005}. The spectral function measured by ARPES is also affected by the Fermi function~\cite{lucy}, matrix elements~\cite{Kurtz_2007} and finite energy and momentum resolution~\cite{conv}. The evaluation of ARPES data based on MDCs, but including finite energy resolution effects, has been also described in Refs.~\cite{Kaminski2005,ingle,Kirkegaard_2005}. Differences between the EDC and the MDC method were also explained for the case of the majority spin band measured on the Fe (110) surface~\cite{Fe011}. Moreover, the MDC and EDC profiles were analyzed in order to fit a phenomenological model describing the spectral function of cuprate superconductors~\cite{edc_model}.

Recently, machine learning techniques have been applied for ARPES data analysis. Some of them are image processing techniques~\cite{deep_neural,convolutional_neural_network}, which can be applied to a two-dimensional ARPES spectrum and enhance its quality (contrast, signal to noise ratio) making the dispersive features more visible, similarly to the curvature method~\cite{curvature} or the minimum gradient method~\cite{minimum_gradient_shen}. Other methods can be also used to extract the self-energy directly from ARPES data~\cite{yamaji2020hidden,bayesian}.


A special approach, which allows to overcome  difficulties in both, MDC and EDC methods has been described in Refs.~\cite{nechaev,li_cuprates,Meevasana2008}. Instead of fitting one-dimensional sections of a spectral function, the two-dimensional function is fitted. The influence of instrumental resolution is taken into account. The number of parameters is significantly reduced in comparison with more standard methods (MDC/EDC analysis) what allows to obtain more robust results. Such an approach has been used for the first time in the analysis of scattering rates of Rashba states ~\cite{nechaev} on Au (111) surface. A linear form of scattering rates and a Rashba-type dispersion have been used in this study. A thorough analysis of the spectral function in the normal and superconducting state of copper-oxide based materials has been also performed using this approach~\cite{li_cuprates}. A specific form of the self-energy has been assumed, but the method has been extended by constraints ensuring consistency between real and imaginary part of the self-energy according to Kramers-Kronig relations. This was possible  for a system with a one band crossing the Fermi level, but it is extremely difficult in other cases. Another variation of this method has been applied for the analysis of the high binding energy ($\sim 0.3$~eV) anomaly in the electronic structure of the Bi$_2$Sr$_2$CuO$_6$ cuprate superconductor~\cite{Meevasana2008}. The bare dispersion has been taken from density functional calculations. Then, both real and imaginary part of the self-energy have been extracted using a two dimensional fit to the ARPES spectrum collected along the nodal direction. Additionally, a momentum dependency of a photoemission matrix element, for a particular experimental geometry, has been determined. The analysis has been done without assuming any particular form of the self-energy, and Kramers-Kronig relations have not been imposed. The influence of instrumental resolution has not been taken into account in the analysis.

\section{Method}

The need of accurate analysis of the spectral function of novel materials has boosted us to develop further two-dimensional fits of ARPES data. Our "All At Once" fitting procedure (AAO) takes into account the influence of both the Fermi function and the instrumental resolution. The spectral function $A$ (eq. \ref{26_1}) is multiplied by the Fermi function. Some form of the dispersion is assumed (e.g. parabolic). Subsequently, it is convoluted with the Gaussian kernel with fixed parameters representing the instrumental resolution. An energy dependent background is added. Then, the entire two dimensional function is fitted to the ARPES spectrum collected along some direction in the momentum space. The scattering rate is obtained directly as a function of energy. For each energy channel $\omega_i$ there is one fit parameter $\Gamma(\omega_i)$.  Our approach does not demand to know 'a priori' a particular form of the scattering rate, in contrast to some previous attempts of two-dimensional fits to ARPES data~\cite{nechaev,li_cuprates}. Additionally,  the AAO analysis is accurate also in the case when instrumental resolution cannot be neglected. This makes our method more general and extremely useful in studies of electronic structure of systems with strongly correlated electrons.

The fitting procedure is sensitive to the relative ratio between energy step of a data $\delta \omega$ and energy resolution $\Delta \omega$. Namely for $\delta \omega < \Delta \omega$, the derived scattering rate starts to oscillate. To overcome this problem we introduced constraints in our fitting program. We assumed the following general form of constraints:
\begin{equation}|\Gamma_i - \hat{\Gamma}_i|<\varepsilon \Gamma_i,\end{equation} where $\varepsilon$ is a small dimensionless number and $\hat{\Gamma}_i$ is a quantity defined by a formula:
\begin{equation}\hat{\Gamma}_i = \sum\limits_{j=-n, j\neq 0}^{n}\Gamma_{i+j}w_j, \end{equation} coefficients $w_j$ are weights coefficients.

We checked that the best fitting results are achieved for $w_j$  of the simplest form, i.e., for $w_j$=const. We aimed on development of some universal constraints, which can be used for any input data. However, we did not succeed, because the proper choice of constraints depends on the scattering rate slope. Therefore, one needs to check carefully that the obtained scattering rates are not distorted by using the constraints.
Here, we performed many fits to the same data set using different constraints (i.e., different combinations of $n$, $\varepsilon$ parameters) to ensure that results are really stable. We also analyzed the correlation matrix between fit coefficients. 

For many systems the spectral function of the form (\ref{26_1}) is modulated by the energy dependent amplitude which is proportional to the renormalization function $Z(\omega)$. The explicit form can be unknown in many cases. Two approaches can be used in order to take into account this effect. Firstly, one can assume that $Z(\omega)$ can be modeled by a polynomial in the considered energy range. Secondly, one can introduce the renormalization function as an independent parameter for each energy channel. However, similarly to the scattering rate, oscillations will appear if one considers data with energy step less than the energy resolution. Additional constraints for $Z(\omega)$ should be used then.  We would like to admit that we do not discuss matrix elements and final states effects~\cite{Kurtz_2007} in the scope of this paper.

In the subsequent sections we analyze the spectral function for different cases: linear band, Dirac cone, hole pocket, Bogoliubov dispersion, and "kink" anomaly due to electron-phonon coupling. We simulated spectral function and then analyzed it with two methods: our new AAO and the standard MDC evaluation method. Moreover, experimental data of K$_{0.4}$Ba$_{0.6}$Fe$_2$As$_2$ are analyzed.   
 
All numerical calculations were performed using Igor Pro (\textit{Wavemetrics}, \href{https://www.wavemetrics.com/}{https://www.wavemetrics.com/}) software. The Levenberg-Marquardt~\cite{levenberg,marquardt} algorithm implemented in Igor Pro was used for non-linear least squares data fitting.

\section{Results and discussion}
\subsection{Linear band crossing the Fermi level}
As a  first example we chose a one-band system with a linear dispersion close to the Fermi level:
\begin{equation}\varepsilon^{\star}(k) = \varepsilon_0 + v k,\label{eq:disp1}\end{equation} where $\varepsilon_0$ is a band energy at momentum $k=0$ and $v$ stands for the band velocity. One branch of the dispersion ($k\ge 0$) has been used in order to avoid complications. Furthermore, Fermi-liquid like scattering rate $\Gamma(\omega)$ is used:
\begin{equation}\Gamma(\omega)=\Gamma_0 + \lambda \omega^2, \end{equation} where $\lambda$ is a coupling constant. Here, the constant term $\Gamma_0$ accounts for the contribution of elastic scattering by defects as well as for the thermal broadening $\Gamma_{th}\sim k_B T$. The values of used parameters are listed in Table~\ref{tab:par1}. We also assume a temperature equal to $T=50$~K. Energy and momentum resolution equal to $\Delta E=12$~meV and $\Delta k=0.022$ 1/\AA $ $ are taken. Such a resolution can be achieved easily in modern ARPES experimental setups, especially those at synchrotron radiation facilities. A background is not added into the simulated spectral function, because we would like to focus here on scattering rates and keep the situation as simple as possible. 

\begin{figure*}
\centering\includegraphics[width=0.9\linewidth]{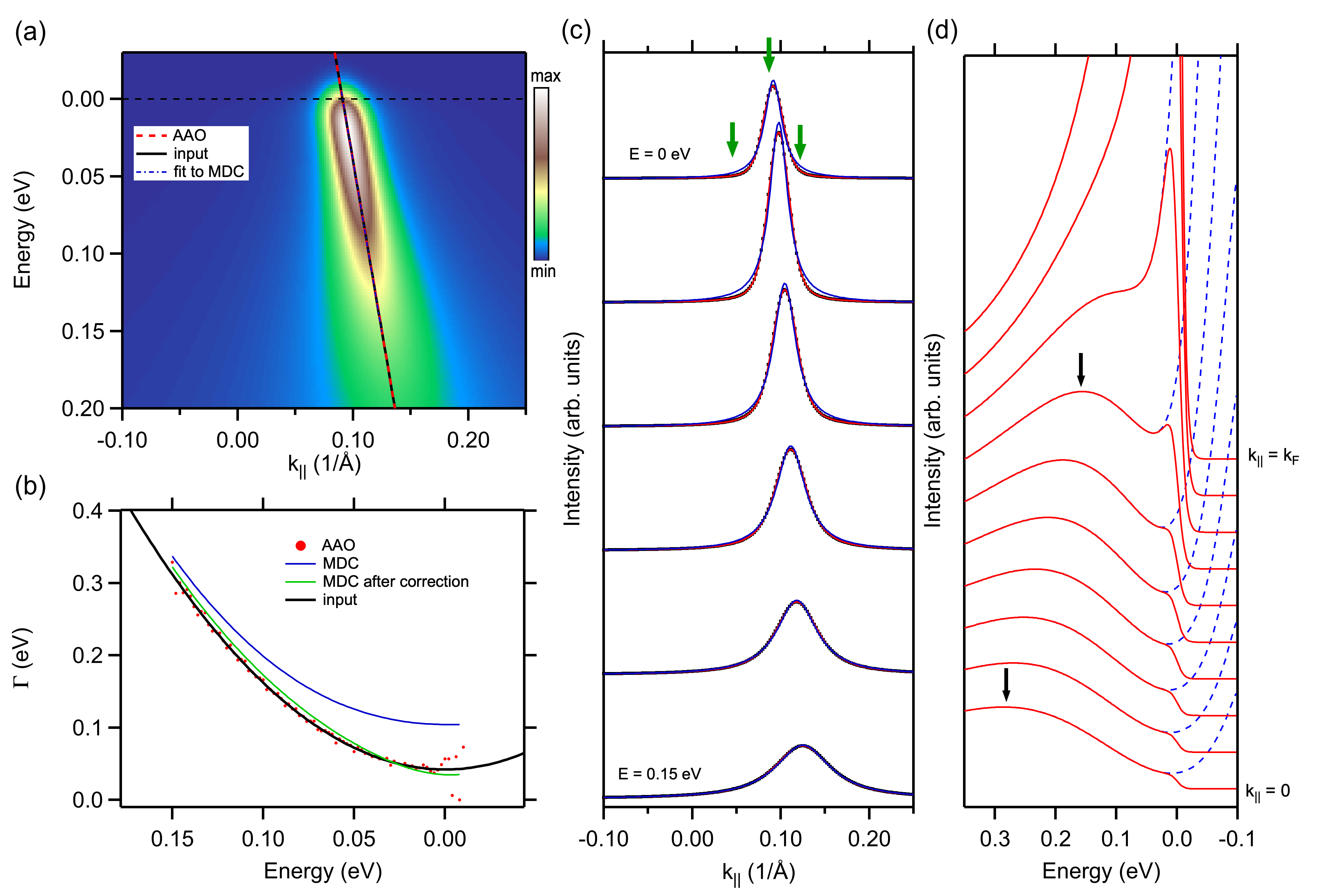}
\caption{The spectral function simulated for the band with a linear dispersion and Fermi liquid like scattering rates. (a) Energy-momentum distribution map. The band dispersion is denoted by the solid black line. Red dashed line stands for the dispersion determined from the AAO fit, while blue dashed-dotted line indicates dispersion obtained from the MDC evaluation. (b) The assumed input Fermi liquid scattering rates (solid black line) are compared with those obtained by the AAO fit (red dots) and the MDC evaluation (blue line). The green line shows MDC evaluation results after correction for finite energy and momentum resolution. (c) Waterfall plot (MDCs) showing comparison between the input data (black dots), the AAO fit (red lines) and the MDC evaluation (blue lines). Green arrows point to the k-region showing strong deviation between different evaluation methods. (d) EDCs corresponding to the spectral function shown in (a) between 0 and $k_F\approx 0.09$ 1/\AA. Black arrows indicate maxima in the spectrum  appearing far from $\varepsilon=\varepsilon^{\star}(k)$ (the "back dispersion" effect). Blue dashed lines show the spectral function without influence of the Fermi function.}
\label{fig:jeden}
\end{figure*}

We fit the simulated data using two approaches: the new AAO fitting procedure, and the standard MDC evaluation (each MDC fitted with a single Lorentzian peak). The obtained results are presented in Fig.~\ref{fig:jeden}. Panel (a) shows a simulated ARPES energy-momentum distribution (EMD) map together with the dispersion (black solid line). The dispersion resulting from the AAO fit (red dashed line) and from the MDC evaluation (blue dashed-dotted line) are also shown in the panel. It is impossible to find any difference between these curves, what is also reflected in the similar band velocities (see Table~\ref{tab:par1}) derived from these methods. The remaining  fit parameters are also collected in Table~\ref{tab:par1}. Fig.~\ref{fig:jeden} shows the scattering rates derived from both fitting methods (AAO: red dots, MDC: blue line) in comparison with input FL scattering rates (black line). The AAO derived scattering rates reproduces the input data almost perfectly well. Only above the Fermi level, some fluctuations appear. This is expected, because above $E_F$ the spectral weight is decreasing rapidly. Thus, a unique determination of $\Gamma(\omega)$ is almost infeasible for $\omega>>k_B T$. It is apparent from Fig.~\ref{fig:jeden} that the MDC evaluation overestimates scattering rates. In this case, the departure from the correct values is more significant close to the~$E_F$. The deviation decreases well below the Fermi level. This effect can be easily explained by analyzing the MDC waterfall plot presented in Fig.~\ref{fig:jeden}~(c). It is apparent that both, MDCs obtained from the MDC evaluation (blue lines) and those obtained from the AAO fit (red lines) are in perfect agreement with the input spectrum (black dots) for $\omega=0.15$~eV. This is because at this energy, the instrumental broadening can be neglected in comparison with the lifetime broadening, and the resulting MDC still can be described by a Lorentzian function (as expected for a linear band). However, close to $E_F$, $\Gamma$ becomes much smaller and one cannot neglect the instrumental resolution. Therefore, the Lorentzian peak cannot accurately describe the data, what is highlighted by green arrows in Fig.~\ref{fig:jeden}~c. However, the position of the maximum is still well reproduced, so the dispersion is still well described close to $E_F$, as evidenced in~Fig.~\ref{fig:jeden}~a.

We tried to correct the MDC derived scattering rates by subtraction of instrumental broadening. In order to do so we treat all spectral lines as Gaussian functions. As visible in Fig.~\ref{fig:jeden}~c, the corrected results (green curve) agree well with the assumed scattering rates.


\begin{table}
\centering
\caption{Comparison between input parameters used in the spectral function simulation for data presented in Fig~\ref{fig:jeden} and results from MDC and AAO fits. MDC: $\Gamma_0$ and $\lambda$ obtained from corrected $\Gamma(\omega)$.\\}

\begin{tabular}{clllll}
 &$\varepsilon_0$ (eV)&$v$ (eV \AA)&$\lambda$ (1/eV)&$\Gamma_0$ (meV)&$Z$ \\
\hline
Input&-0.4&4.4&12.00&42.199&1000 \\
MDC& -0.406(1)&4.456(8)&12.6(1)&44(1)&-  \\
AAO&-0.39977&4.39769&12.03(8)&42.2(8)&999.521\\
\hline
\label{tab:par1}
\end{tabular}
\end{table}

In Fig.~\ref{fig:jeden}~d we show selected EDCs derived from the EMD map shown in panel (a). These curves correspond to $k$ values between 0 and $k_F\approx 0.09$~1/\AA. We also plot curves corresponding to the spectral function without influence of the Fermi function (blue dashed lines). There are some broad maxima in the spectral function, highlighted by black arrows, which do not correspond to $\varepsilon^{\star}(k)$. It can be explained as follows: for non-interacting electrons, the spectral function displays maxima at $(k, \varepsilon(k))$. In case of a system of interacting electrons, the charge carrier lifetime is not infinite and this leads to spectrum broadening, which is usually energy-dependent. In consequence, part of the spectral weight stemming from the dispersive band $\varepsilon^{\star}(k)$ can be transferred to other locations in $(k,\omega)$ phase space. This can be visible as presence of additional maxima in the spectral function, and it is sometimes called as the "back-dispersion"~\cite{HUFNER1999191}. This is caused by the spectral function above the Fermi level in case of larger lifetime broadening. This effect can lead to confusions and makes that the analysis of spectra of strongly correlated systems  is highly nontrivial. 


\subsection{Dirac cone dispersion}
\begin{figure}
\centering\includegraphics[width=0.9\linewidth]{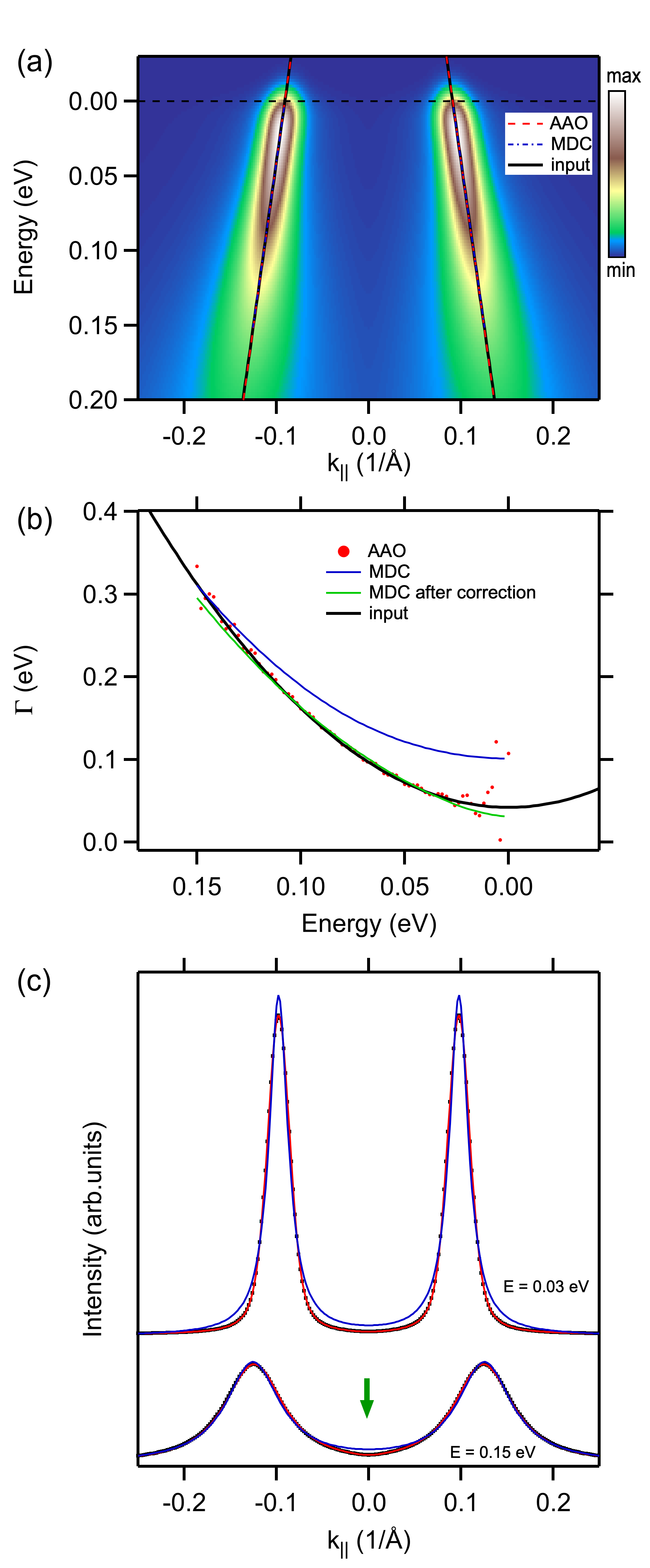}
\caption{The spectral function simulated for a Dirac dispersion and Fermi liquid scattering rates. (a) EMD map with the input dispersion (black solid line) and the dispersion obtained from the MDC evaluation (blue dashed-dotted line), and from the AAO fit (dashed red line). (b) The assumed input Fermi liquid scattering rates (solid black line) are compared with those obtained by the AAO fit (red dots) and the MDC evaluation (blue line). The green line shows the MDC evaluation results after correction for finite energy and momentum resolution. (c) Exemplary MDCs:  from input data (black dots), AAO fit (red lines) and MDC evaluation (blue lines).  The green arrow indicates a cusp in the spectral function at $k_{||}=0$. }
\label{fig:dwa}
\end{figure}

Materials with a dispersion relation described by a Dirac cone in a momentum space, such as e.g. graphene or topological insulators, attracted recently particular attention. We simulated the spectral function of the Dirac material in the same way as described in previous section with the only difference that instead of using eq. \ref{eq:disp1}, we describe the dispersion with:
\begin{equation}\varepsilon^{\star}(k)=\varepsilon_0 + v |k|,\end{equation} where $\varepsilon_0$ is an energy corresponding to the Dirac point and $v$ is a slope of the band. We assume that the Dirac point is well above $E_F$ ($\varepsilon_0=-0.4$~eV). Thus, the spectral contribution from the upper part of the cone can be neglected in the region of our interest. We assume Fermi liquid-like scattering rates. Similarly as for a single branch of a linear band, here both the MDC and the AAO method determine the dispersion relation correctly (Fig.~\ref{fig:dwa}~a). The relation between the input scattering rate and those determined by fitting with the two methods (Fig.~\ref{fig:dwa}~b) is also the same as in the  previous section. We notice that the  AAO derived scattering rates become noisy close to $E_F$. The so called ''back-dispersion'' effect can be also found in the simulated spectral function in this case. However, some difference is visible in the shape of the spectral function close to $k\approx 0$ (cf. Fig.~\ref{fig:dwa}~c). Namely, the spectral function calculated for a Dirac dispersion displays a cusp at $k=0$ (first derivative of a dispersion relation is not a continuous function of $k$ close to this point). Such a feature cannot be described properly by a combination of two Lorentzian functions used in the MDC evaluation, what is apparent in Fig.~\ref{fig:dwa}~c (green arrow points to the position of the cusp). Of course, in the experimental data, such a cusp is blurred due to the influence of instrumental resolution. Parameters obtained from fits to the data presented in Fig.~\ref{fig:dwa} are provided in Table~\ref{tab:par2}.

\begin{table}
\centering
\caption{Comparison between input parameters used in the spectral function simulation for data presented in Fig~\ref{fig:dwa} and the results from MDC and AAO fits. MDC: $\Gamma_0$ and $\lambda$ estimated from corrected $\Gamma(\omega)$.\\}

\begin{tabular}{clllll}
 &$\varepsilon_0$ (eV)&$v$ (eV \AA)&$\lambda$ (1/eV)&$\Gamma_0$ (meV)&$Z$ \\
\hline
Input&-0.4&4.4&12.00&42.199&1000 \\
MDC&-0.396&4.36&11.645&42&-  \\
AAO&-0.404&4.44&12.2&42(1)&1008.7\\
\hline
\label{tab:par2}
\end{tabular}
\end{table}

\subsection{Parabolic hole pocket}
\begin{figure*}
\centering\includegraphics[width=0.8\linewidth]{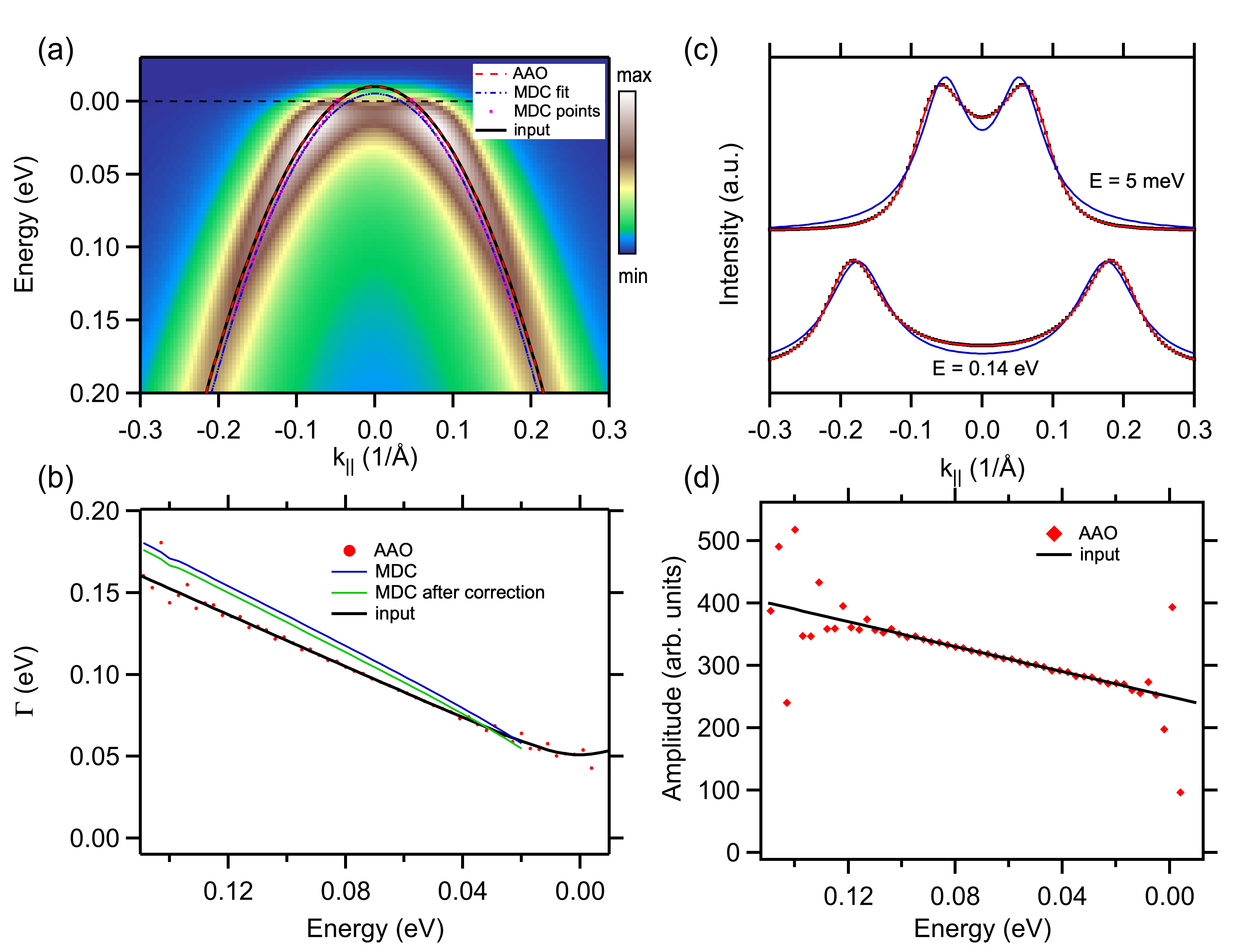}
\caption{The spectral function simulated for a parabolic dispersion and marginal Fermi liquid scattering rates. (a) EMD map with the input dispersion (black solid line), the dispersion obtained from the MDC evaluation (magenta dots), and from the AAO fit (dashed red line). The MDC dispersion was fitted with a parabola (blue dashed-dotted line). (b) The assumed input non-Fermi liquid scattering rates (solid black line) are compared with those obtained from the AAO fit (red dots) and the MDC evaluation (blue line). The green line shows the MDC evaluation results after correction for finite energy and momentum resolution. (c) Exemplary MDCs: input data (black dots), AAO evaluation (red lines), and MDC evaluation (blue lines). (d) The energy-dependent amplitude obtained from the AAO fit (red markers) is compared with the input amplitude (black line).}
\label{fig:trzy}
\end{figure*}

In this case, the band dispersion is given by:
\begin{equation}\varepsilon^{\star}(k) = \varepsilon_0 + a k^2.\label{eq:parabolic}\end{equation}
Here, we model a situation with a hole band with its top located 10~meV above the Fermi level. We assume a lifetime broadening consistent with the marginal Fermi liquid theory (MFL) \cite{Varma1989,varma1}, which has been observed experimentally for many materials with strong electronic correlations (such as cuprates, iron-based superconductors). The total spectral width is given by~\cite{Varma1989,varma1}:
 \begin{equation}\Gamma(\omega)=\lambda \sqrt{\omega^2 + \left({\pi} k_B T\right)^2} + \Gamma_{el},\label{eq:mfl}\end{equation} $\lambda$ is a dimensionless coupling constant and $\Gamma_{el}$ is an elastic contribution to the scattering rate. Here, we approximate $\Gamma_{el}$ by a constant and we use a parabola for the renormalized dispersion. We present results corresponding to $\lambda=0.8$,  which is close but still below the Planckian limit (i.e., $\lambda=1$) \cite{planckian_1, planckian_2}. Our results are shown in Fig.~\ref{fig:trzy}. We assume here, that the renormalization function can be approximated by a linear function in energy, i.e., there is an energy-dependent amplitude $Z=a_0+a_1 \omega$. Such a energy dependence of the renormalization function $Z(E)$, is visible in ARPES spectral function of materials and provides important information about the correlation effects~\cite{linkage}.
  
The differences between AAO and MDC methods are visible in the derived dispersion (Fig.~\ref{fig:trzy}~a). The AAO fit (red dashed line) recovers the dispersion almost perfectly, while the MDC points (magenta dots) do not agree with the input dispersion (black line). The MDC points can be described by a parabola (blue dashed-dotted line) for $\omega>0.02$~eV, but it is apparent that the top of the band obtained from extrapolation is underestimated. Close to $E_F$ the slope of the MDC derived band is completely incorrect. Further differences can be noticed in the derived scattering rates (Fig.~\ref{fig:trzy}~b) and in shapes of MDC lines (Fig.~\ref{fig:trzy}~c). The AAO fit gives correct scattering rates below $E_F$. Above $E_F$ some oscillations are visible. The slope of $\Gamma(E)$ obtained from the MDC evaluation is too large at higher binding energies. Close to $E_F$, the curvature of MDC derived $\Gamma(E)$ dependency is not correct. The energy-dependent amplitude obtained from the AAO fit is visible in Fig.~\ref{fig:trzy}~a (red markers) and compared with the input amplitude (black line). Good agreement between both is visible between 0.12 eV and 0.006 eV. Pronounced oscillations are apparent close to $E_F$  and for $\omega> 0.12$~eV in the AAO result. They appear, because of edge effects, which are difficult to avoid when calculating the convolution, and because of a rapid decrease of intensity caused by the Fermi function.  The comparison between the two evaluation methods is summarized in Table~\ref{tab:par3}. 
 
\begin{table}
\centering
\caption{Comparison between input parameters used in the spectral function simulation for data presented in Fig~\ref{fig:trzy} and results from MDC and AAO fits. MDC: $\Gamma_0$ and $\lambda$ estimated from corrected $\Gamma(\omega)$.\\}

\begin{tabular}{cllllll}
 &$\varepsilon_0$ (meV)&$a$ (eV \AA$^2$)&$\lambda$&$\Gamma_0$ (meV)&$a_0$&$a_1$ \\
\hline
Input&-10&4.5&0.8&40&250&1000 \\
MDC&-6.13&4.73&0.922&39.7(2)&-&-  \\
AAO&-9.96&4.496&0.797&40.1(2)&-249.8&1004.2\\
\hline
\label{tab:par3}
\end{tabular}
\end{table}

\subsection{K$_{0.4}$Ba$_{0.6}$Fe$_2$As$_2$ data - inner hole pocket around $\Gamma$ point}
\begin{figure}
\centering\includegraphics[width=0.9\linewidth]{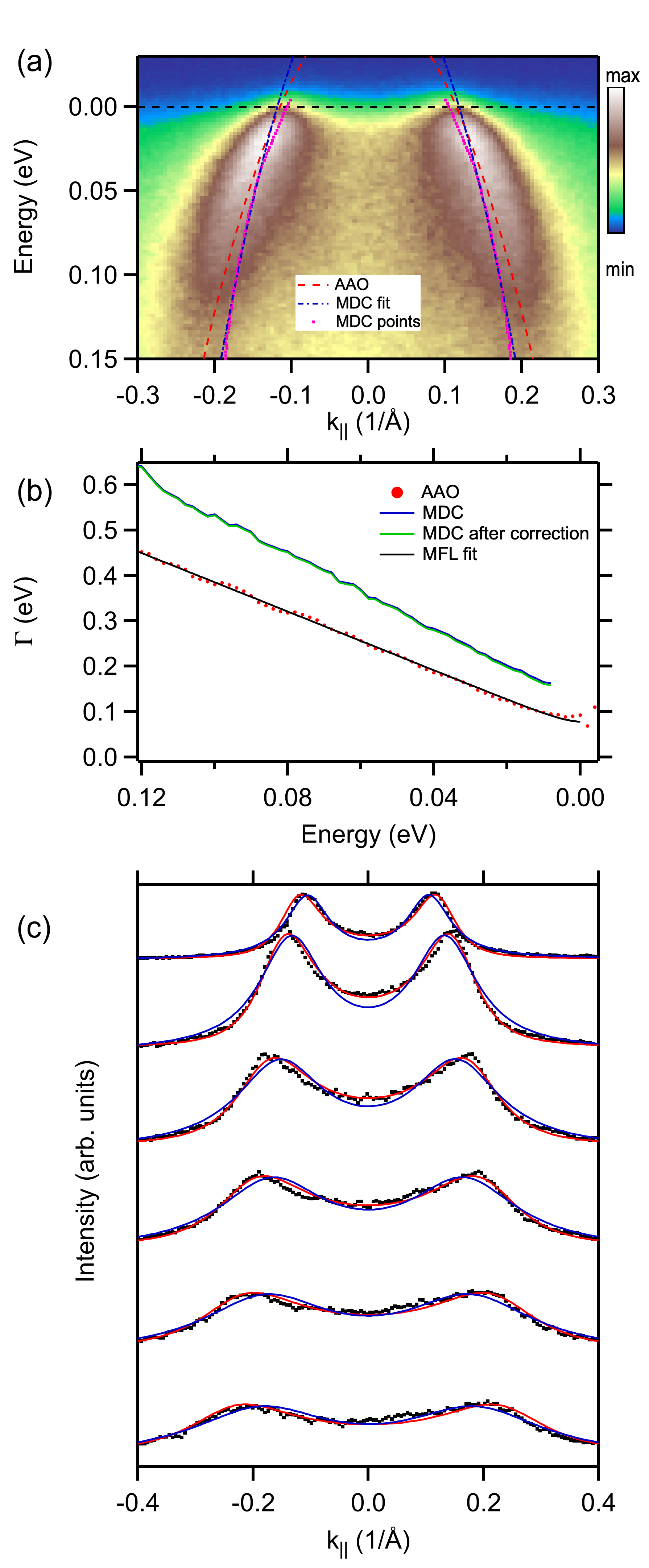}
\caption{K$_{0.4}$Ba$_{0.6}$Fe$_2$As$_2$ ARPES data for inner hole pocket around the $\Gamma$ point measured at T=50 K. (a) EMD map together with a dispersion obtained from the MDC evaluation (magenta dots) and from the AAO fit (dashed red line). The MDC dispersion was fitted with a parabola (blue dashed-dotted line). (b) Scattering rates obtained from the AAO fit (red dots) and the MDC evaluation (blue line). The green line shows MDC evaluation results after correction for finite energy and momentum resolution. The black line is a marginal Fermi liquid fit to the AAO result. (c) Exemplary MDCs: experimental data (black dots), AAO evaluation (red lines) and MDC evaluation (blue lines). Energy range: between 0.15 and 0 eV. }
\label{fig:cztery}
\end{figure}

Electronic correlations are considered as a crucial factor for superconductivity in iron-based pnictides and chalcogenides. They can lead to the strong renormalization of the band structure which can be studied by ARPES. Here, compared to the previous section, we now present an analysis of experimental ARPES data of  K$_{0.4}$Ba$_{0.6}$Fe$_2$As$_2$ (optimally doped BaFe$_2$As$_2$, $T_c\approx35$~ K) collected at T=50~K, partially already published in~\cite{linkage}. The data collected along $\Gamma$-M direction are shown in Fig.~\ref{fig:cztery}~a together with dispersions obtained by AAO (dashed red line) and MDC (magenta dots) evaluations. The AAO fit was performed assuming a linear in energy and a linear in momentum background. We also used  a quadratic in energy amplitude ($a=a_0+a_1 \omega +
a_2 \omega^2$) which multiples spectral function in order to achieve good agreement with the experimental data. Such an additional modulation of the ARPES signal is an effect of the energy dependent renormalization function $Z$  which is consistent with the MFL theory. In case of the AAO evaluation we checked that a parabolic form of $\varepsilon^{\star}(k)$ works the best. The MDC derived points can be fitted with a parabola between 0.15 and -0.03 eV (blue dashed-dotted line). A small difference between the MDC and the AAO derived dispersions is visible.

We fitted the AAO derived scattering rates with the dependency predicted by MFL theory (eq.~\ref{eq:mfl}). The elastic contribution to spectral broadening was described by the relation~\cite{kevan}:
\begin{equation}\Gamma_{el}(\omega)=\Gamma(0) \frac{v(\omega)}{v(0)},\end{equation} where $v(\omega)$ is a group velocity, $\Gamma(0)$ and $v(0)$ denote scattering rate and velocity at the Fermi level, respectively. We obtained a value for the coupling constant $\lambda=3.03$, which is well above the Planckian limit (i.e., $\lambda=1$). Such a large value suggests that a normal state of this material is highly incoherent. This topic and the relation between scattering rates and transition temperature $T_c$ are discussed in detail elsewhere \cite{linkage}. The elastic scattering at $E_F$ amounts to $\Gamma_{el}(0)=47(1)$~meV. It is apparent from Fig.~\ref{fig:cztery}~b that the slope of the MDC derived scattering rate (blue curve) is larger compared to the AAO derived scattering rate and cannot be reduced by a correction for finite experimental resolution (green line). The coupling constant derived from the MDC evaluation reaches $\lambda=3.9$. All parameters obtained using AAO and MDC methods are collected in Table~\ref{tab:par4}.

\begin{table}
\centering
\caption{Parameters obtained from MDC and AAO fits of K$_{0.4}$Ba$_{0.6}$Fe$_2$As$_2$ data (Fig.~\ref{fig:cztery}). MDC: $\Gamma_0$ and $\lambda$ obtained from corrected $\Gamma(\omega)$.\\}
\begin{tabular}{clllll}
 &$\varepsilon_0$ (meV)&$a$ (eV \AA$^2$)&$\lambda$&$\Gamma_{el}(0)$ (meV) \\
\hline
MDC&92.9&-6.62&3.93(3)&95(2)  \\
AAO&60.3&-4.56&3.03(2)&47(1)\\
\hline
\label{tab:par4}
\end{tabular}
\end{table}

Good agreement between the spectral function modeled with the aid of the AAO approach and the ARPES data is evidenced by a waterfall plot (Fig.~\ref{fig:cztery}~c). A significant discrepancy between the experimental data and the MDC double Lorentzian model close to $E_F$ in the region of $k_{||}\approx 0$ is also visible.

\subsection{Spectral function in superconducting (BCS) state}
\begin{figure}
\centering\includegraphics[width=0.9\linewidth]{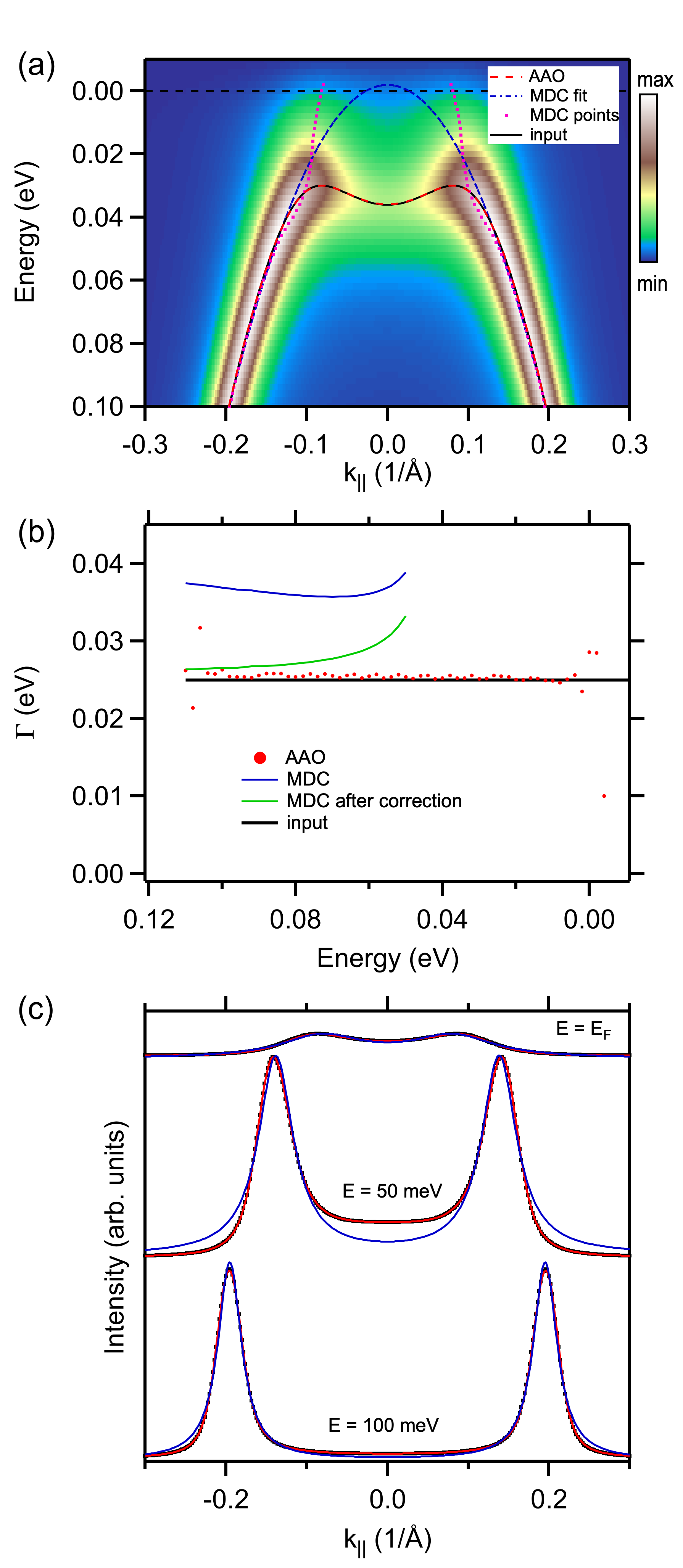}
\caption{The spectral function simulated for the BCS dispersion and constant scattering rates. (a) EMD map  with the input dispersion (black solid line) and the dispersion obtained from the MDC evaluation (magenta dots) and from the AAO fit (dashed red line).  The MDC dispersion was fitted with a parabola (blue dashed-dotted line) (b) Input scattering rates (solid black line) are compared with those obtained by the AAO fit (red dots) and the MDC evaluation (blue line). The green line shows MDC evaluation results after correction for finite energy and momentum resolution. (c) Exemplary MDCs: input data (black dots), AAO evaluation (red lines) and MDC evaluation (blue lines). }
\label{fig:piatka}
\end{figure}

Properties of conventional superconductors can be usually explained by the Bardeen-Cooper-Schrieffer (BCS) theory. Their electronic structure can be described by the Bogoliubov dispersion close to the Fermi level:
\begin{equation}E(k) = \sqrt{\varepsilon^2(k)+\Delta^2},\end{equation} where $\varepsilon(k)$ is a bare band dispersion and $\Delta$ is the superconducting gap parameter. Here, we simulated the spectral function for a BCS state (Fig.~\ref{fig:piatka}) using an approach published before \cite{rinott}. The spectral function can be written as:
\begin{equation}A(k,\omega)=\frac{u_k^2}{\pi} \frac{\frac{\Gamma}{2}}{(\omega-E(k))^2+(\frac{\Gamma}{2})^2}+\frac{v_k^2}{\pi} \frac{\frac{\Gamma}{2}}{(\omega+E(k))^2+(\frac{\Gamma}{2})^2},\end{equation} where $u_k$ and $v_k$ are parameters describing superconducting wave function:
\begin{equation}v_k^2=\frac{1}{2}\left(1-\frac{\varepsilon(k)}{E(k)}\right)\end{equation}
\begin{equation}u_k^2 = 1 - v_k^2. \end{equation}
We assumed that the bare band is a parabolic hole pocket (cf. eq.~\ref{eq:parabolic}), and used constant scattering rates $\Gamma$. All parameters are summarized in Table~\ref{tab:par5}. It is apparent from Fig.~\ref{fig:piatka} that the AAO method reproduces perfectly dispersion in the superconducting state. The MDC derived points cannot describe the shape of the band close to $E_F$, where the superconducting gap is present. This is because of the "back-bending"  of the band due to electron-hole mixing. On the other hand, the MDC derived dispersion between 0.1 eV and 0.05 eV can be well described by a parabolic function. The parabolic fit in this range was used to determine scattering rates which are shown Fig.~\ref{fig:piatka}~b (blue line) together with the AAO result (red dots) and the input scattering rate (black line). The MDC scattering rates are too big and are not constant as a function of energy. The correction for a finite instrumental resolution reduces values of $\Gamma$, but its shape (see green curve in Fig.~\ref{fig:piatka}~b) is still far from being constant. The correct values of $\Gamma$ are approached for $|\omega|>>\Delta$. The strange behavior of the MDC derived scattering rates can be understood looking at the waterfall plot (Fig.~\ref{fig:piatka}~c) which shows that the spectral function at a given energy $\omega$ cannot be correctly modeled by the sum of Lorentzian peaks unless $|\omega|>>\Delta$. We conclude that the MDC evaluation itself cannot be used for the analysis of a BCS-like electronic structure. However, it can be combined with other methods, e.g. with fitting a Dynes function~\cite{dynes_gap} to the density of states derived from ARPES. 


\begin{table}
\centering
\caption{Comparison between input parameters used in the spectral function simulation for data presented in Fig~\ref{fig:piatka} and the results from MDC and AAO fits.\\}

\begin{tabular}{clllll}
 &$\varepsilon_0$ (meV)&$a$ (eV \AA$^2$)&$\Delta$ (meV)&$\Gamma$ (meV)&$Z$ \\
\hline
Input&-20&3&30&25&1000 \\
MDC&-1.86&2.653&-&-&-  \\
AAO&-19.79&2.988&30.03&-&1000.08\\
\hline
\label{tab:par5}
\end{tabular}
\end{table}

\subsection{"Kink" anomaly due to electron-phonon coupling}
The interaction between electrons and the lattice dynamics can alter significantly the band structure (see e.g. Refs.~\cite{Bianchi2010,shin_Rb}). Here, we consider the simplest possible case, i.e., coupling to a single phonon mode localized at some energy $\Omega$. We limit ourself to low temperatures (T=5 K). The real part of the self energy can be calculated using~\cite{fonony}:
\begin{equation}\Re\Sigma(\omega)=\frac{\Omega\lambda}{2}\ln{\frac{(\omega+\Omega)^2+\gamma^2}{(\omega-\Omega)^2+\gamma^2}},\end{equation} where $\lambda$ denotes a coupling constant, and $\gamma$ is a small real number that ensures finite value of $\Re\Sigma$ at energy equal to $\Omega$. We chose $\gamma = 10^{-3}$~eV.  The imaginary part of the self-energy operator $\Im\Sigma$ is usually modeled  by the step function: 
\begin{equation}\Im\Sigma(\omega) = - \frac{\lambda \Omega \pi}{2}\quad \textnormal{for}\quad | \omega |> \Omega; \quad \Im\Sigma=0\quad \textnormal{otherwise}.\label{eq:im_sigma}\end{equation} 
However, the sharp jump in $\Im\Sigma$ at $\omega = \Omega $ appears only if the renormalization effect in $\Re\Sigma$ goes to infinity at $\Omega$. Therefore, instead of using eq.~\ref{eq:im_sigma} , we modeled $\Im\Sigma$ using a smooth function:
\begin{equation}\Im\Sigma(\omega) =-\frac{\lambda \Omega \pi}{2}(1+\tanh\left(\frac{\omega-\Omega}{w}\right)),
\label{eq:imsig}\end{equation} where we choose $w=0.01$~eV.

\begin{figure}
\centering\includegraphics[width=0.9\linewidth]{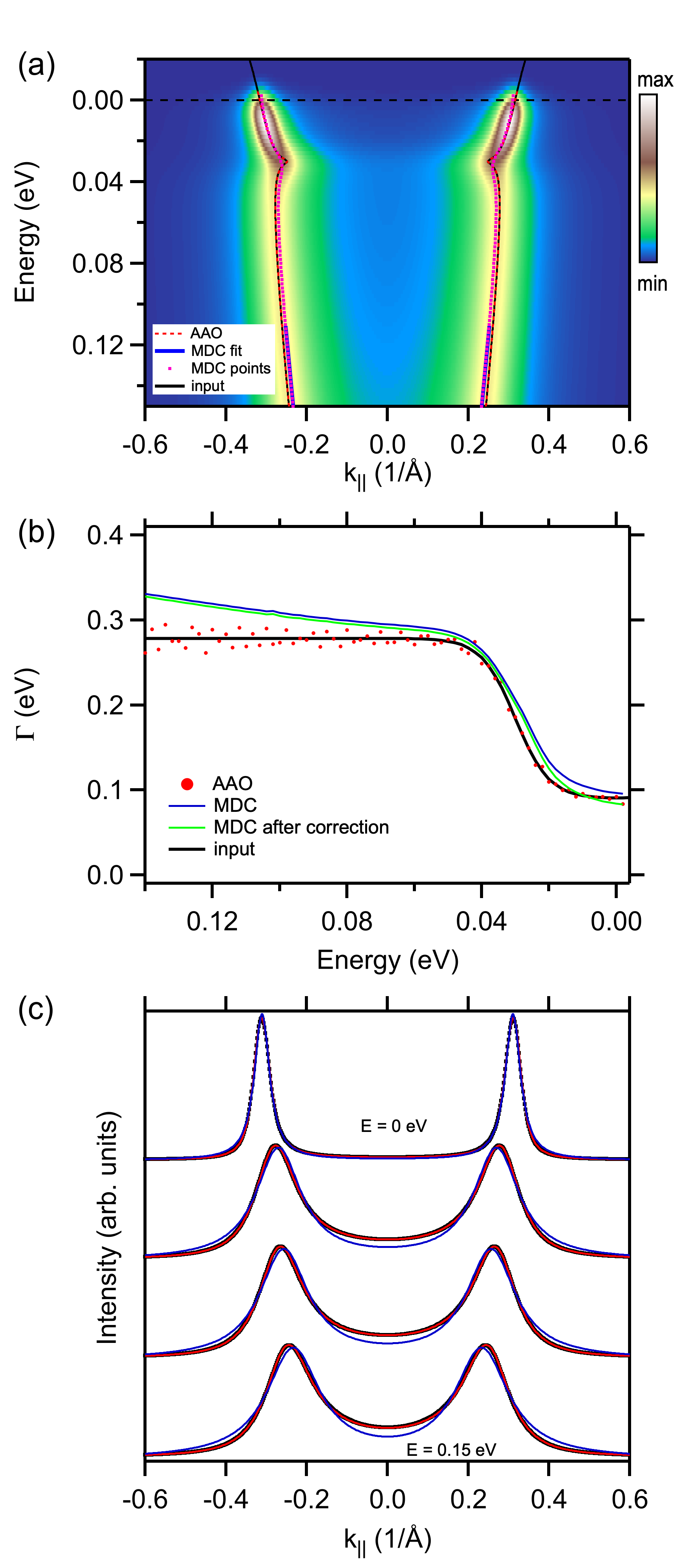}
\caption{The spectral function simulated for a parabolic band with a "kink" due to electron-phonon interaction. (a) EMD map with the input dispersion (black solid line) and the dispersion obtained from the MDC evaluation (magenta dots) and from the AAO fit (dashed red line). The MDC dispersion was fitted with a linear function (blue line) (b) Input scattering rates (solid black line) are compared with those obtained from the AAO fit (red dots) and the MDC evaluation (blue line). The green line shows MDC evaluation results after correction for finite energy and momentum resolution. (c) Exemplary MDCs: input data (black dots), AAO evaluation (red lines) and MDC evaluation (blue lines). }
\label{fig:szesc}
\end{figure}

 The spectral function was simulated using a parabolic bare band and assuming parameters similar to those which can be found in other articles~\cite{Fink2007,Hofmann_2009}. Our results are shown in Fig.~\ref{fig:szesc}. The input dispersion (black line, panel a) is a solution to the equation:
\begin{equation}\omega-\varepsilon(k)-\Re\Sigma(\omega)=0 \end{equation} and it is well reproduced by the AAO fit (red dashed line). The MDC derived dispersion (magenta dots) agrees well with the input for $\omega>\Omega$, while for $\omega<\Omega$ there is a significant difference between them. We fitted a linear function (blue line) to the MDC derived points between 0.15 eV and 0.01 eV, in order to estimate a bare band velocity necessary to calculate the scattering rates. Fig.~\ref{fig:szesc}~b provides a comparison between input (black line), AAO (red markers) and MDC (blue line) derived spectral widths. Correction for finite energy and momentum resolution applied to the MDC result (see the green curve) does not play a role. The overall profile of the input scattering rate is reproduced correctly by the AAO fit, but one should notice that significant fluctuations appear, especially for $|\omega|>\Omega$. On the other hand, the MDC result is stable in the full analyzed energy range. It is also apparent that both methods provide a proper estimation of $\Omega$. Systematic difference between the AAO and the MDC derived scattering rates are present for $|\omega|>\Omega$, in particular in the region where the spectral function is not properly described by a double Lorentzian, what is visible in the waterfall plot (Fig.~\ref{fig:szesc}~c). We compare all fit coefficients in Table~\ref{tab:par6}. 
The coupling constant $\lambda$ and phonon energy $\Omega$ were obtained from fit of formula~(\ref{eq:imsig}) to the MDC  derived scattering rate corrected for a finite resolution.

\begin{table}
\centering
\caption{Comparison between input parameters used in the spectral function simulation for the data presented in Fig~\ref{fig:szesc} and results from MDC and AAO fits.\\}

\begin{tabular}{cllllll}
 &$\varepsilon_0$ (eV)&$a$ (eV \AA$^2$)&$\lambda$&$\Omega$ (meV)&$\gamma$ (meV)&$Z$ \\
\hline
Input&0.4&-4&1&30&1&1000 \\
MDC&0.387(1)&-4.36(1)&1.18&28.2&-&-  \\
AAO&0.399&-3.994&0.998&29.98&0.99&998.49\\
\hline
\label{tab:par6}
\end{tabular}
\end{table}

\section{Conclusions}
In this article we discussed several possible approaches which can be used for the analysis of ARPES data. We briefly reviewed traditional methods focusing attention on advantages and drawbacks of them. A new method (AAO) has been proposed. It allows for accurate evaluation of ARPES data. This was demonstrated for several types of band structures realized in real materials, i.e., for a Dirac dispersion, a parabolic hole pocket, for the Bogoliubov dispersion in a superconducting state, and for a band with a "kink" anomaly due to electron-phonon coupling. The data corresponding to the inner hole pocket around the $\Gamma$ point measured for K$_{0.4}$Ba$_{0.6}$Fe$_2$As$_2$ were analyzed with the new, proposed method. This method uses the exact formula for the spectral function, taking into account many-body effects. Additionally, obtained results are free from the influence of the instrumental resolution. The advantage of AAO approach is particularly visible for ARPES data  close close to the top/bottom of a hole/electron pocket and close to the edge of the superconducting gap. In particular, we emphasize that with our new method, also the energy dependence of parameters, such as the scattering rate, the dispersion (not shown) or the renormalization function~\cite{linkage} can be determined without knowing or modeling their energy dependence before. We believe that our proposition can be useful for further studies of novel quantum materials.











\section{ ACKNOWLEDGMENTS}
We thank Emile Rienks  (HZB) and A. Garrison Linn (CU Boulder) for helpful discussions.
\bibliographystyle{apsrev4-1}
\bibliography{ARPESmany}

\begin{thebibliography}{46}%
\makeatletter
\providecommand \@ifxundefined [1]{%
 \@ifx{#1\undefined}
}%
\providecommand \@ifnum [1]{%
 \ifnum #1\expandafter \@firstoftwo
 \else \expandafter \@secondoftwo
 \fi
}%
\providecommand \@ifx [1]{%
 \ifx #1\expandafter \@firstoftwo
 \else \expandafter \@secondoftwo
 \fi
}%
\providecommand \natexlab [1]{#1}%
\providecommand \enquote  [1]{``#1''}%
\providecommand \bibnamefont  [1]{#1}%
\providecommand \bibfnamefont [1]{#1}%
\providecommand \citenamefont [1]{#1}%
\providecommand \href@noop [0]{\@secondoftwo}%
\providecommand \href [0]{\begingroup \@sanitize@url \@href}%
\providecommand \@href[1]{\@@startlink{#1}\@@href}%
\providecommand \@@href[1]{\endgroup#1\@@endlink}%
\providecommand \@sanitize@url [0]{\catcode `\\12\catcode `\$12\catcode
  `\&12\catcode `\#12\catcode `\^12\catcode `\_12\catcode `\%12\relax}%
\providecommand \@@startlink[1]{}%
\providecommand \@@endlink[0]{}%
\providecommand \url  [0]{\begingroup\@sanitize@url \@url }%
\providecommand \@url [1]{\endgroup\@href {#1}{\urlprefix }}%
\providecommand \urlprefix  [0]{URL }%
\providecommand \Eprint [0]{\href }%
\providecommand \doibase [0]{http://dx.doi.org/}%
\providecommand \selectlanguage [0]{\@gobble}%
\providecommand \bibinfo  [0]{\@secondoftwo}%
\providecommand \bibfield  [0]{\@secondoftwo}%
\providecommand \translation [1]{[#1]}%
\providecommand \BibitemOpen [0]{}%
\providecommand \bibitemStop [0]{}%
\providecommand \bibitemNoStop [0]{.\EOS\space}%
\providecommand \EOS [0]{\spacefactor3000\relax}%
\providecommand \BibitemShut  [1]{\csname bibitem#1\endcsname}%
\let\auto@bib@innerbib\@empty
\bibitem [{\citenamefont {Sobota}\ \emph {et~al.}(2021)\citenamefont {Sobota},
  \citenamefont {He},\ and\ \citenamefont {Shen}}]{sobota_review}%
  \BibitemOpen
  \bibfield  {author} {\bibinfo {author} {\bibfnamefont {J.~A.}\ \bibnamefont
  {Sobota}}, \bibinfo {author} {\bibfnamefont {Y.}~\bibnamefont {He}}, \ and\
  \bibinfo {author} {\bibfnamefont {Z.-X.}\ \bibnamefont {Shen}},\ }\href
  {\doibase 10.1103/RevModPhys.93.025006} {\bibfield  {journal} {\bibinfo
  {journal} {Rev. Mod. Phys.}\ }\textbf {\bibinfo {volume} {93}},\ \bibinfo
  {pages} {025006} (\bibinfo {year} {2021})}\BibitemShut {NoStop}%
\bibitem [{\citenamefont {Damascelli}\ \emph {et~al.}(2003)\citenamefont
  {Damascelli}, \citenamefont {Hussain},\ and\ \citenamefont
  {Shen}}]{Damascelli2003}%
  \BibitemOpen
  \bibfield  {author} {\bibinfo {author} {\bibfnamefont {A.}~\bibnamefont
  {Damascelli}}, \bibinfo {author} {\bibfnamefont {Z.}~\bibnamefont {Hussain}},
  \ and\ \bibinfo {author} {\bibfnamefont {Z.-X.}\ \bibnamefont {Shen}},\
  }\href {\doibase 10.1103/RevModPhys.75.473} {\bibfield  {journal} {\bibinfo
  {journal} {Rev. Mod. Phys.}\ }\textbf {\bibinfo {volume} {75}},\ \bibinfo
  {pages} {473} (\bibinfo {year} {2003})}\BibitemShut {NoStop}%
\bibitem [{\citenamefont {Mahan}(2000)}]{Mahan2000}%
  \BibitemOpen
  \bibfield  {author} {\bibinfo {author} {\bibfnamefont {G.~D.}\ \bibnamefont
  {Mahan}},\ }\href@noop {} {\emph {\bibinfo {title} {Many-Particle Physics}}}\
  (\bibinfo  {publisher} {Kluwer Academic/Plenum Publishers, New York},\
  \bibinfo {year} {2000})\BibitemShut {NoStop}%
\bibitem [{\citenamefont {Ortenzi}\ \emph {et~al.}(2009)\citenamefont
  {Ortenzi}, \citenamefont {Cappelluti}, \citenamefont {Benfatto},\ and\
  \citenamefont {Pietronero}}]{ortenzi}%
  \BibitemOpen
  \bibfield  {author} {\bibinfo {author} {\bibfnamefont {L.}~\bibnamefont
  {Ortenzi}}, \bibinfo {author} {\bibfnamefont {E.}~\bibnamefont {Cappelluti}},
  \bibinfo {author} {\bibfnamefont {L.}~\bibnamefont {Benfatto}}, \ and\
  \bibinfo {author} {\bibfnamefont {L.}~\bibnamefont {Pietronero}},\ }\href
  {\doibase 10.1103/PhysRevLett.103.046404} {\bibfield  {journal} {\bibinfo
  {journal} {Phys. Rev. Lett.}\ }\textbf {\bibinfo {volume} {103}},\ \bibinfo
  {pages} {046404} (\bibinfo {year} {2009})}\BibitemShut {NoStop}%
\bibitem [{\citenamefont {LaShell}\ \emph {et~al.}(2000)\citenamefont
  {LaShell}, \citenamefont {Jensen},\ and\ \citenamefont
  {Balasubramanian}}]{LaShell2000}%
  \BibitemOpen
  \bibfield  {author} {\bibinfo {author} {\bibfnamefont {S.}~\bibnamefont
  {LaShell}}, \bibinfo {author} {\bibfnamefont {E.}~\bibnamefont {Jensen}}, \
  and\ \bibinfo {author} {\bibfnamefont {T.}~\bibnamefont {Balasubramanian}},\
  }\href {\doibase 10.1103/PhysRevB.61.2371} {\bibfield  {journal} {\bibinfo
  {journal} {Phys. Rev. B}\ }\textbf {\bibinfo {volume} {61}},\ \bibinfo
  {pages} {2371} (\bibinfo {year} {2000})}\BibitemShut {NoStop}%
\bibitem [{\citenamefont {Valla}\ \emph
  {et~al.}(1999{\natexlab{a}})\citenamefont {Valla}, \citenamefont {Fedorov},
  \citenamefont {Johnson},\ and\ \citenamefont {Hulbert}}]{Valla1999}%
  \BibitemOpen
  \bibfield  {author} {\bibinfo {author} {\bibfnamefont {T.}~\bibnamefont
  {Valla}}, \bibinfo {author} {\bibfnamefont {A.~V.}\ \bibnamefont {Fedorov}},
  \bibinfo {author} {\bibfnamefont {P.~D.}\ \bibnamefont {Johnson}}, \ and\
  \bibinfo {author} {\bibfnamefont {S.~L.}\ \bibnamefont {Hulbert}},\ }\href
  {\doibase 10.1103/PhysRevLett.83.2085} {\bibfield  {journal} {\bibinfo
  {journal} {Phys. Rev. Lett.}\ }\textbf {\bibinfo {volume} {83}},\ \bibinfo
  {pages} {2085} (\bibinfo {year} {1999}{\natexlab{a}})}\BibitemShut {NoStop}%
\bibitem [{\citenamefont {Higashiguchi}\ \emph {et~al.}(2007)\citenamefont
  {Higashiguchi}, \citenamefont {Shimada}, \citenamefont {Arita}, \citenamefont
  {Miura}, \citenamefont {Tobita}, \citenamefont {Cui}, \citenamefont {Aiura},
  \citenamefont {Namatame},\ and\ \citenamefont
  {Taniguchi}}]{ni_scattering_rates}%
  \BibitemOpen
  \bibfield  {author} {\bibinfo {author} {\bibfnamefont {M.}~\bibnamefont
  {Higashiguchi}}, \bibinfo {author} {\bibfnamefont {K.}~\bibnamefont
  {Shimada}}, \bibinfo {author} {\bibfnamefont {M.}~\bibnamefont {Arita}},
  \bibinfo {author} {\bibfnamefont {Y.}~\bibnamefont {Miura}}, \bibinfo
  {author} {\bibfnamefont {N.}~\bibnamefont {Tobita}}, \bibinfo {author}
  {\bibfnamefont {X.}~\bibnamefont {Cui}}, \bibinfo {author} {\bibfnamefont
  {Y.}~\bibnamefont {Aiura}}, \bibinfo {author} {\bibfnamefont
  {H.}~\bibnamefont {Namatame}}, \ and\ \bibinfo {author} {\bibfnamefont
  {M.}~\bibnamefont {Taniguchi}},\ }\href {\doibase
  https://doi.org/10.1016/j.susc.2007.04.136} {\bibfield  {journal} {\bibinfo
  {journal} {Surface Science}\ }\textbf {\bibinfo {volume} {601}},\ \bibinfo
  {pages} {4005} (\bibinfo {year} {2007})},\ \bibinfo {note}
  {eCOSS-24}\BibitemShut {NoStop}%
\bibitem [{\citenamefont {Hayashi}\ \emph {et~al.}(2013)\citenamefont
  {Hayashi}, \citenamefont {Shimada}, \citenamefont {Jiang}, \citenamefont
  {Iwasawa}, \citenamefont {Aiura}, \citenamefont {Oguchi}, \citenamefont
  {Namatame},\ and\ \citenamefont {Taniguchi}}]{pallad}%
  \BibitemOpen
  \bibfield  {author} {\bibinfo {author} {\bibfnamefont {H.}~\bibnamefont
  {Hayashi}}, \bibinfo {author} {\bibfnamefont {K.}~\bibnamefont {Shimada}},
  \bibinfo {author} {\bibfnamefont {J.}~\bibnamefont {Jiang}}, \bibinfo
  {author} {\bibfnamefont {H.}~\bibnamefont {Iwasawa}}, \bibinfo {author}
  {\bibfnamefont {Y.}~\bibnamefont {Aiura}}, \bibinfo {author} {\bibfnamefont
  {T.}~\bibnamefont {Oguchi}}, \bibinfo {author} {\bibfnamefont
  {H.}~\bibnamefont {Namatame}}, \ and\ \bibinfo {author} {\bibfnamefont
  {M.}~\bibnamefont {Taniguchi}},\ }\href {\doibase 10.1103/PhysRevB.87.035140}
  {\bibfield  {journal} {\bibinfo  {journal} {Phys. Rev. B}\ }\textbf {\bibinfo
  {volume} {87}},\ \bibinfo {pages} {035140} (\bibinfo {year}
  {2013})}\BibitemShut {NoStop}%
\bibitem [{\citenamefont {Jiang}\ \emph {et~al.}(2009)\citenamefont {Jiang},
  \citenamefont {Higashiguchi}, \citenamefont {Tobida}, \citenamefont {Tanaka},
  \citenamefont {Fukuda}, \citenamefont {Hayashi}, \citenamefont {Shimada},
  \citenamefont {Namatame},\ and\ \citenamefont {Taniguchi}}]{al_kink}%
  \BibitemOpen
  \bibfield  {author} {\bibinfo {author} {\bibfnamefont {J.}~\bibnamefont
  {Jiang}}, \bibinfo {author} {\bibfnamefont {M.}~\bibnamefont {Higashiguchi}},
  \bibinfo {author} {\bibfnamefont {N.}~\bibnamefont {Tobida}}, \bibinfo
  {author} {\bibfnamefont {K.}~\bibnamefont {Tanaka}}, \bibinfo {author}
  {\bibfnamefont {S.}~\bibnamefont {Fukuda}}, \bibinfo {author} {\bibfnamefont
  {H.}~\bibnamefont {Hayashi}}, \bibinfo {author} {\bibfnamefont
  {K.}~\bibnamefont {Shimada}}, \bibinfo {author} {\bibfnamefont
  {H.}~\bibnamefont {Namatame}}, \ and\ \bibinfo {author} {\bibfnamefont
  {M.}~\bibnamefont {Taniguchi}},\ }\href {\doibase 10.1380/ejssnt.2009.57}
  {\bibfield  {journal} {\bibinfo  {journal} {e-Journal of Surface Science and
  Nanotechnology}\ }\textbf {\bibinfo {volume} {7}},\ \bibinfo {pages} {57}
  (\bibinfo {year} {2009})}\BibitemShut {NoStop}%
\bibitem [{\citenamefont {Reinert}\ \emph {et~al.}(2004)\citenamefont
  {Reinert}, \citenamefont {Eltner}, \citenamefont {Nicolay}, \citenamefont
  {Forster}, \citenamefont {Schmidt},\ and\ \citenamefont
  {Hüfner}}]{reinert_Pb_Cu}%
  \BibitemOpen
  \bibfield  {author} {\bibinfo {author} {\bibfnamefont {F.}~\bibnamefont
  {Reinert}}, \bibinfo {author} {\bibfnamefont {B.}~\bibnamefont {Eltner}},
  \bibinfo {author} {\bibfnamefont {G.}~\bibnamefont {Nicolay}}, \bibinfo
  {author} {\bibfnamefont {F.}~\bibnamefont {Forster}}, \bibinfo {author}
  {\bibfnamefont {S.}~\bibnamefont {Schmidt}}, \ and\ \bibinfo {author}
  {\bibfnamefont {S.}~\bibnamefont {Hüfner}},\ }\href {\doibase
  https://doi.org/10.1016/j.physb.2004.06.013} {\bibfield  {journal} {\bibinfo
  {journal} {Physica B: Condensed Matter}\ }\textbf {\bibinfo {volume} {351}},\
  \bibinfo {pages} {229} (\bibinfo {year} {2004})},\ \bibinfo {note}
  {proceedings of The International Symposium on Synchrotron Radiation Research
  for Spin and Electronic States in d and f Electron Systems}\BibitemShut
  {NoStop}%
\bibitem [{\citenamefont {Matzdorf}(1998)}]{MATZDORF}%
  \BibitemOpen
  \bibfield  {author} {\bibinfo {author} {\bibfnamefont {R.}~\bibnamefont
  {Matzdorf}},\ }\href {\doibase https://doi.org/10.1016/S0167-5729(97)00013-7}
  {\bibfield  {journal} {\bibinfo  {journal} {Surface Science Reports}\
  }\textbf {\bibinfo {volume} {30}},\ \bibinfo {pages} {153} (\bibinfo {year}
  {1998})}\BibitemShut {NoStop}%
\bibitem [{\citenamefont {Valla}\ \emph
  {et~al.}(1999{\natexlab{b}})\citenamefont {Valla}, \citenamefont {Fedorov},
  \citenamefont {Johnson},\ and\ \citenamefont {Hulbert}}]{Valla1999b}%
  \BibitemOpen
  \bibfield  {author} {\bibinfo {author} {\bibfnamefont {T.}~\bibnamefont
  {Valla}}, \bibinfo {author} {\bibfnamefont {A.~V.}\ \bibnamefont {Fedorov}},
  \bibinfo {author} {\bibfnamefont {P.~D.}\ \bibnamefont {Johnson}}, \ and\
  \bibinfo {author} {\bibfnamefont {S.~L.}\ \bibnamefont {Hulbert}},\ }\href
  {http://link.aps.org/doi/10.1103/PhysRevLett.83.2085} {\bibfield  {journal}
  {\bibinfo  {journal} {Phys. Rev. Lett.}\ }\textbf {\bibinfo {volume} {83}},\
  \bibinfo {pages} {2085} (\bibinfo {year} {1999}{\natexlab{b}})}\BibitemShut
  {NoStop}%
\bibitem [{\citenamefont {Kordyuk}\ \emph {et~al.}(2005)\citenamefont
  {Kordyuk}, \citenamefont {Borisenko}, \citenamefont {Koitzsch}, \citenamefont
  {Fink}, \citenamefont {Knupfer},\ and\ \citenamefont {Berger}}]{Kordyuk2005}%
  \BibitemOpen
  \bibfield  {author} {\bibinfo {author} {\bibfnamefont {A.~A.}\ \bibnamefont
  {Kordyuk}}, \bibinfo {author} {\bibfnamefont {S.~V.}\ \bibnamefont
  {Borisenko}}, \bibinfo {author} {\bibfnamefont {A.}~\bibnamefont {Koitzsch}},
  \bibinfo {author} {\bibfnamefont {J.}~\bibnamefont {Fink}}, \bibinfo {author}
  {\bibfnamefont {M.}~\bibnamefont {Knupfer}}, \ and\ \bibinfo {author}
  {\bibfnamefont {H.}~\bibnamefont {Berger}},\ }\href {\doibase
  10.1103/PhysRevB.71.214513} {\bibfield  {journal} {\bibinfo  {journal} {Phys.
  Rev. B}\ }\textbf {\bibinfo {volume} {71}},\ \bibinfo {pages} {214513}
  (\bibinfo {year} {2005})}\BibitemShut {NoStop}%
\bibitem [{\citenamefont {Rameau}\ \emph {et~al.}(2010)\citenamefont {Rameau},
  \citenamefont {Yang},\ and\ \citenamefont {Johnson}}]{lucy}%
  \BibitemOpen
  \bibfield  {author} {\bibinfo {author} {\bibfnamefont {J.}~\bibnamefont
  {Rameau}}, \bibinfo {author} {\bibfnamefont {H.-B.}\ \bibnamefont {Yang}}, \
  and\ \bibinfo {author} {\bibfnamefont {P.}~\bibnamefont {Johnson}},\ }\href
  {\doibase https://doi.org/10.1016/j.elspec.2010.05.025} {\bibfield  {journal}
  {\bibinfo  {journal} {Journal of Electron Spectroscopy and Related
  Phenomena}\ }\textbf {\bibinfo {volume} {181}},\ \bibinfo {pages} {35}
  (\bibinfo {year} {2010})},\ \bibinfo {note} {proceedings of International
  Workshop on Strong Correlations and Angle-Resolved Photoemission Spectroscopy
  2009}\BibitemShut {NoStop}%
\bibitem [{\citenamefont {Kurtz}\ \emph {et~al.}(2007)\citenamefont {Kurtz},
  \citenamefont {Browne},\ and\ \citenamefont {Mankey}}]{Kurtz_2007}%
  \BibitemOpen
  \bibfield  {author} {\bibinfo {author} {\bibfnamefont {R.~L.}\ \bibnamefont
  {Kurtz}}, \bibinfo {author} {\bibfnamefont {D.~A.}\ \bibnamefont {Browne}}, \
  and\ \bibinfo {author} {\bibfnamefont {G.~J.}\ \bibnamefont {Mankey}},\
  }\href {\doibase 10.1088/0953-8984/19/35/355001} {\bibfield  {journal}
  {\bibinfo  {journal} {Journal of Physics: Condensed Matter}\ }\textbf
  {\bibinfo {volume} {19}},\ \bibinfo {pages} {355001} (\bibinfo {year}
  {2007})}\BibitemShut {NoStop}%
\bibitem [{\citenamefont {Helander}\ \emph {et~al.}(2011)\citenamefont
  {Helander}, \citenamefont {Greiner}, \citenamefont {Wang},\ and\
  \citenamefont {Lu}}]{conv}%
  \BibitemOpen
  \bibfield  {author} {\bibinfo {author} {\bibfnamefont {M.~G.}\ \bibnamefont
  {Helander}}, \bibinfo {author} {\bibfnamefont {M.~T.}\ \bibnamefont
  {Greiner}}, \bibinfo {author} {\bibfnamefont {Z.~B.}\ \bibnamefont {Wang}}, \
  and\ \bibinfo {author} {\bibfnamefont {Z.~H.}\ \bibnamefont {Lu}},\ }\href
  {\doibase 10.1063/1.3642659} {\bibfield  {journal} {\bibinfo  {journal}
  {Review of Scientific Instruments}\ }\textbf {\bibinfo {volume} {82}},\
  \bibinfo {pages} {096107} (\bibinfo {year} {2011})}\BibitemShut {NoStop}%
\bibitem [{\citenamefont {Kaminski}\ and\ \citenamefont
  {Fretwell}(2005)}]{Kaminski2005}%
  \BibitemOpen
  \bibfield  {author} {\bibinfo {author} {\bibfnamefont {A.}~\bibnamefont
  {Kaminski}}\ and\ \bibinfo {author} {\bibfnamefont {H.~M.}\ \bibnamefont
  {Fretwell}},\ }\href {http://dx.doi.org/10.1088/1367-2630/7/1/098} {\bibfield
   {journal} {\bibinfo  {journal} {New Journal of Physics}\ }\textbf {\bibinfo
  {volume} {7}},\ \bibinfo {pages} {98} (\bibinfo {year} {2005})}\BibitemShut
  {NoStop}%
\bibitem [{\citenamefont {Ingle}\ \emph {et~al.}(2005)\citenamefont {Ingle},
  \citenamefont {Shen}, \citenamefont {Baumberger}, \citenamefont {Meevasana},
  \citenamefont {Lu}, \citenamefont {Shen}, \citenamefont {Damascelli},
  \citenamefont {Nakatsuji}, \citenamefont {Mao}, \citenamefont {Maeno},
  \citenamefont {Kimura},\ and\ \citenamefont {Tokura}}]{ingle}%
  \BibitemOpen
  \bibfield  {author} {\bibinfo {author} {\bibfnamefont {N.~J.~C.}\
  \bibnamefont {Ingle}}, \bibinfo {author} {\bibfnamefont {K.~M.}\ \bibnamefont
  {Shen}}, \bibinfo {author} {\bibfnamefont {F.}~\bibnamefont {Baumberger}},
  \bibinfo {author} {\bibfnamefont {W.}~\bibnamefont {Meevasana}}, \bibinfo
  {author} {\bibfnamefont {D.~H.}\ \bibnamefont {Lu}}, \bibinfo {author}
  {\bibfnamefont {Z.-X.}\ \bibnamefont {Shen}}, \bibinfo {author}
  {\bibfnamefont {A.}~\bibnamefont {Damascelli}}, \bibinfo {author}
  {\bibfnamefont {S.}~\bibnamefont {Nakatsuji}}, \bibinfo {author}
  {\bibfnamefont {Z.~Q.}\ \bibnamefont {Mao}}, \bibinfo {author} {\bibfnamefont
  {Y.}~\bibnamefont {Maeno}}, \bibinfo {author} {\bibfnamefont
  {T.}~\bibnamefont {Kimura}}, \ and\ \bibinfo {author} {\bibfnamefont
  {Y.}~\bibnamefont {Tokura}},\ }\href {\doibase 10.1103/PhysRevB.72.205114}
  {\bibfield  {journal} {\bibinfo  {journal} {Phys. Rev. B}\ }\textbf {\bibinfo
  {volume} {72}},\ \bibinfo {pages} {205114} (\bibinfo {year}
  {2005})}\BibitemShut {NoStop}%
\bibitem [{\citenamefont {Kirkegaard}\ \emph {et~al.}(2005)\citenamefont
  {Kirkegaard}, \citenamefont {Kim},\ and\ \citenamefont
  {Hofmann}}]{Kirkegaard_2005}%
  \BibitemOpen
  \bibfield  {author} {\bibinfo {author} {\bibfnamefont {C.}~\bibnamefont
  {Kirkegaard}}, \bibinfo {author} {\bibfnamefont {T.~K.}\ \bibnamefont {Kim}},
  \ and\ \bibinfo {author} {\bibfnamefont {P.}~\bibnamefont {Hofmann}},\ }\href
  {\doibase 10.1088/1367-2630/7/1/099} {\bibfield  {journal} {\bibinfo
  {journal} {New Journal of Physics}\ }\textbf {\bibinfo {volume} {7}},\
  \bibinfo {pages} {99} (\bibinfo {year} {2005})}\BibitemShut {NoStop}%
\bibitem [{\citenamefont {Cui}\ \emph {et~al.}(2010)\citenamefont {Cui},
  \citenamefont {Shimada}, \citenamefont {Sakisaka}, \citenamefont {Kato},
  \citenamefont {Hoesch}, \citenamefont {Oguchi}, \citenamefont {Aiura},
  \citenamefont {Namatame},\ and\ \citenamefont {Taniguchi}}]{Fe011}%
  \BibitemOpen
  \bibfield  {author} {\bibinfo {author} {\bibfnamefont {X.~Y.}\ \bibnamefont
  {Cui}}, \bibinfo {author} {\bibfnamefont {K.}~\bibnamefont {Shimada}},
  \bibinfo {author} {\bibfnamefont {Y.}~\bibnamefont {Sakisaka}}, \bibinfo
  {author} {\bibfnamefont {H.}~\bibnamefont {Kato}}, \bibinfo {author}
  {\bibfnamefont {M.}~\bibnamefont {Hoesch}}, \bibinfo {author} {\bibfnamefont
  {T.}~\bibnamefont {Oguchi}}, \bibinfo {author} {\bibfnamefont
  {Y.}~\bibnamefont {Aiura}}, \bibinfo {author} {\bibfnamefont
  {H.}~\bibnamefont {Namatame}}, \ and\ \bibinfo {author} {\bibfnamefont
  {M.}~\bibnamefont {Taniguchi}},\ }\href {\doibase 10.1103/PhysRevB.82.195132}
  {\bibfield  {journal} {\bibinfo  {journal} {Phys. Rev. B}\ }\textbf {\bibinfo
  {volume} {82}},\ \bibinfo {pages} {195132} (\bibinfo {year}
  {2010})}\BibitemShut {NoStop}%
\bibitem [{\citenamefont {Matsuyama}\ and\ \citenamefont
  {Gweon}(2013)}]{edc_model}%
  \BibitemOpen
  \bibfield  {author} {\bibinfo {author} {\bibfnamefont {K.}~\bibnamefont
  {Matsuyama}}\ and\ \bibinfo {author} {\bibfnamefont {G.-H.}\ \bibnamefont
  {Gweon}},\ }\href {\doibase 10.1103/PhysRevLett.111.246401} {\bibfield
  {journal} {\bibinfo  {journal} {Phys. Rev. Lett.}\ }\textbf {\bibinfo
  {volume} {111}},\ \bibinfo {pages} {246401} (\bibinfo {year}
  {2013})}\BibitemShut {NoStop}%
\bibitem [{\citenamefont {Kim}\ \emph {et~al.}(2021)\citenamefont {Kim},
  \citenamefont {Oh}, \citenamefont {Huh}, \citenamefont {Song}, \citenamefont
  {Jeong}, \citenamefont {Kwon}, \citenamefont {Kim}, \citenamefont {Kim},
  \citenamefont {Ryu}, \citenamefont {Jung}, \citenamefont {Kyung},
  \citenamefont {Sohn}, \citenamefont {Lee}, \citenamefont {Hyun},
  \citenamefont {Lee}, \citenamefont {Kim},\ and\ \citenamefont
  {Kim}}]{deep_neural}%
  \BibitemOpen
  \bibfield  {author} {\bibinfo {author} {\bibfnamefont {Y.}~\bibnamefont
  {Kim}}, \bibinfo {author} {\bibfnamefont {D.}~\bibnamefont {Oh}}, \bibinfo
  {author} {\bibfnamefont {S.}~\bibnamefont {Huh}}, \bibinfo {author}
  {\bibfnamefont {D.}~\bibnamefont {Song}}, \bibinfo {author} {\bibfnamefont
  {S.}~\bibnamefont {Jeong}}, \bibinfo {author} {\bibfnamefont
  {J.}~\bibnamefont {Kwon}}, \bibinfo {author} {\bibfnamefont {M.}~\bibnamefont
  {Kim}}, \bibinfo {author} {\bibfnamefont {D.}~\bibnamefont {Kim}}, \bibinfo
  {author} {\bibfnamefont {H.}~\bibnamefont {Ryu}}, \bibinfo {author}
  {\bibfnamefont {J.}~\bibnamefont {Jung}}, \bibinfo {author} {\bibfnamefont
  {W.}~\bibnamefont {Kyung}}, \bibinfo {author} {\bibfnamefont
  {B.}~\bibnamefont {Sohn}}, \bibinfo {author} {\bibfnamefont {S.}~\bibnamefont
  {Lee}}, \bibinfo {author} {\bibfnamefont {J.}~\bibnamefont {Hyun}}, \bibinfo
  {author} {\bibfnamefont {Y.}~\bibnamefont {Lee}}, \bibinfo {author}
  {\bibfnamefont {Y.}~\bibnamefont {Kim}}, \ and\ \bibinfo {author}
  {\bibfnamefont {C.}~\bibnamefont {Kim}},\ }\href {\doibase 10.1063/5.0054920}
  {\bibfield  {journal} {\bibinfo  {journal} {Review of Scientific
  Instruments}\ }\textbf {\bibinfo {volume} {92}},\ \bibinfo {pages} {073901}
  (\bibinfo {year} {2021})}\BibitemShut {NoStop}%
\bibitem [{\citenamefont {Peng}\ \emph {et~al.}(2020)\citenamefont {Peng},
  \citenamefont {Gao}, \citenamefont {He}, \citenamefont {Li}, \citenamefont
  {Ji}, \citenamefont {Liu}, \citenamefont {Ekahana}, \citenamefont {Pei},
  \citenamefont {Liu}, \citenamefont {Shen},\ and\ \citenamefont
  {Chen}}]{convolutional_neural_network}%
  \BibitemOpen
  \bibfield  {author} {\bibinfo {author} {\bibfnamefont {H.}~\bibnamefont
  {Peng}}, \bibinfo {author} {\bibfnamefont {X.}~\bibnamefont {Gao}}, \bibinfo
  {author} {\bibfnamefont {Y.}~\bibnamefont {He}}, \bibinfo {author}
  {\bibfnamefont {Y.}~\bibnamefont {Li}}, \bibinfo {author} {\bibfnamefont
  {Y.}~\bibnamefont {Ji}}, \bibinfo {author} {\bibfnamefont {C.}~\bibnamefont
  {Liu}}, \bibinfo {author} {\bibfnamefont {S.~A.}\ \bibnamefont {Ekahana}},
  \bibinfo {author} {\bibfnamefont {D.}~\bibnamefont {Pei}}, \bibinfo {author}
  {\bibfnamefont {Z.}~\bibnamefont {Liu}}, \bibinfo {author} {\bibfnamefont
  {Z.}~\bibnamefont {Shen}}, \ and\ \bibinfo {author} {\bibfnamefont
  {Y.}~\bibnamefont {Chen}},\ }\href {\doibase 10.1063/1.5132586} {\bibfield
  {journal} {\bibinfo  {journal} {Review of Scientific Instruments}\ }\textbf
  {\bibinfo {volume} {91}},\ \bibinfo {pages} {033905} (\bibinfo {year}
  {2020})}\BibitemShut {NoStop}%
\bibitem [{\citenamefont {Zhang}\ \emph {et~al.}(2011)\citenamefont {Zhang},
  \citenamefont {Richard}, \citenamefont {Qian}, \citenamefont {Xu},
  \citenamefont {Dai},\ and\ \citenamefont {Ding}}]{curvature}%
  \BibitemOpen
  \bibfield  {author} {\bibinfo {author} {\bibfnamefont {P.}~\bibnamefont
  {Zhang}}, \bibinfo {author} {\bibfnamefont {P.}~\bibnamefont {Richard}},
  \bibinfo {author} {\bibfnamefont {T.}~\bibnamefont {Qian}}, \bibinfo {author}
  {\bibfnamefont {Y.-M.}\ \bibnamefont {Xu}}, \bibinfo {author} {\bibfnamefont
  {X.}~\bibnamefont {Dai}}, \ and\ \bibinfo {author} {\bibfnamefont
  {H.}~\bibnamefont {Ding}},\ }\href {\doibase 10.1063/1.3585113} {\bibfield
  {journal} {\bibinfo  {journal} {Review of Scientific Instruments}\ }\textbf
  {\bibinfo {volume} {82}},\ \bibinfo {pages} {043712} (\bibinfo {year}
  {2011})}\BibitemShut {NoStop}%
\bibitem [{\citenamefont {He}\ \emph {et~al.}(2017)\citenamefont {He},
  \citenamefont {Wang},\ and\ \citenamefont {Shen}}]{minimum_gradient_shen}%
  \BibitemOpen
  \bibfield  {author} {\bibinfo {author} {\bibfnamefont {Y.}~\bibnamefont
  {He}}, \bibinfo {author} {\bibfnamefont {Y.}~\bibnamefont {Wang}}, \ and\
  \bibinfo {author} {\bibfnamefont {Z.-X.}\ \bibnamefont {Shen}},\ }\href
  {\doibase 10.1063/1.4993919} {\bibfield  {journal} {\bibinfo  {journal}
  {Review of Scientific Instruments}\ }\textbf {\bibinfo {volume} {88}},\
  \bibinfo {pages} {073903} (\bibinfo {year} {2017})}\BibitemShut {NoStop}%
\bibitem [{\citenamefont {Yamaji}\ \emph {et~al.}(2020)\citenamefont {Yamaji},
  \citenamefont {Yoshida}, \citenamefont {Fujimori},\ and\ \citenamefont
  {Imada}}]{yamaji2020hidden}%
  \BibitemOpen
  \bibfield  {author} {\bibinfo {author} {\bibfnamefont {Y.}~\bibnamefont
  {Yamaji}}, \bibinfo {author} {\bibfnamefont {T.}~\bibnamefont {Yoshida}},
  \bibinfo {author} {\bibfnamefont {A.}~\bibnamefont {Fujimori}}, \ and\
  \bibinfo {author} {\bibfnamefont {M.}~\bibnamefont {Imada}},\ }\href@noop {}
  {\enquote {\bibinfo {title} {Hidden self-energies as origin of cuprate
  superconductivity revealed by machine learning},}\ } (\bibinfo {year}
  {2020}),\ \Eprint {http://arxiv.org/abs/1903.08060} {arXiv:1903.08060
  [cond-mat.str-el]} \BibitemShut {NoStop}%
\bibitem [{\citenamefont {Tokuda}\ \emph {et~al.}(2021)\citenamefont {Tokuda},
  \citenamefont {Souma}, \citenamefont {Segawa}, \citenamefont {Takahashi},
  \citenamefont {Ando}, \citenamefont {Nakanishi},\ and\ \citenamefont
  {Sato}}]{bayesian}%
  \BibitemOpen
  \bibfield  {author} {\bibinfo {author} {\bibfnamefont {S.}~\bibnamefont
  {Tokuda}}, \bibinfo {author} {\bibfnamefont {S.}~\bibnamefont {Souma}},
  \bibinfo {author} {\bibfnamefont {K.}~\bibnamefont {Segawa}}, \bibinfo
  {author} {\bibfnamefont {T.}~\bibnamefont {Takahashi}}, \bibinfo {author}
  {\bibfnamefont {Y.}~\bibnamefont {Ando}}, \bibinfo {author} {\bibfnamefont
  {T.}~\bibnamefont {Nakanishi}}, \ and\ \bibinfo {author} {\bibfnamefont
  {T.}~\bibnamefont {Sato}},\ }\href {\doibase 10.1038/s42005-021-00673-6}
  {\bibfield  {journal} {\bibinfo  {journal} {Commun. Phys.}\ }\textbf
  {\bibinfo {volume} {4}} (\bibinfo {year} {2021}),\
  10.1038/s42005-021-00673-6}\BibitemShut {NoStop}%
\bibitem [{\citenamefont {Nechaev}\ \emph {et~al.}(2009)\citenamefont
  {Nechaev}, \citenamefont {Jensen}, \citenamefont {Rienks}, \citenamefont
  {Silkin}, \citenamefont {Echenique}, \citenamefont {Chulkov},\ and\
  \citenamefont {Hofmann}}]{nechaev}%
  \BibitemOpen
  \bibfield  {author} {\bibinfo {author} {\bibfnamefont {I.~A.}\ \bibnamefont
  {Nechaev}}, \bibinfo {author} {\bibfnamefont {M.~F.}\ \bibnamefont {Jensen}},
  \bibinfo {author} {\bibfnamefont {E.~D.~L.}\ \bibnamefont {Rienks}}, \bibinfo
  {author} {\bibfnamefont {V.~M.}\ \bibnamefont {Silkin}}, \bibinfo {author}
  {\bibfnamefont {P.~M.}\ \bibnamefont {Echenique}}, \bibinfo {author}
  {\bibfnamefont {E.~V.}\ \bibnamefont {Chulkov}}, \ and\ \bibinfo {author}
  {\bibfnamefont {P.}~\bibnamefont {Hofmann}},\ }\href {\doibase
  10.1103/PhysRevB.80.113402} {\bibfield  {journal} {\bibinfo  {journal} {Phys.
  Rev. B}\ }\textbf {\bibinfo {volume} {80}},\ \bibinfo {pages} {113402}
  (\bibinfo {year} {2009})}\BibitemShut {NoStop}%
\bibitem [{\citenamefont {Li}\ \emph {et~al.}(2018)\citenamefont {Li},
  \citenamefont {Zhou}, \citenamefont {Parham}, \citenamefont {Reber},
  \citenamefont {Berger}, \citenamefont {Arnold},\ and\ \citenamefont
  {Dessau}}]{li_cuprates}%
  \BibitemOpen
  \bibfield  {author} {\bibinfo {author} {\bibfnamefont {H.}~\bibnamefont
  {Li}}, \bibinfo {author} {\bibfnamefont {X.}~\bibnamefont {Zhou}}, \bibinfo
  {author} {\bibfnamefont {S.}~\bibnamefont {Parham}}, \bibinfo {author}
  {\bibfnamefont {T.~J.}\ \bibnamefont {Reber}}, \bibinfo {author}
  {\bibfnamefont {H.}~\bibnamefont {Berger}}, \bibinfo {author} {\bibfnamefont
  {G.~B.}\ \bibnamefont {Arnold}}, \ and\ \bibinfo {author} {\bibfnamefont
  {D.~S.}\ \bibnamefont {Dessau}},\ }\href {\doibase
  10.1038/s41467-017-02422-2} {\bibfield  {journal} {\bibinfo  {journal} {Nat.
  Commun.}\ }\textbf {\bibinfo {volume} {9}},\ \bibinfo {pages} {26} (\bibinfo
  {year} {2018})}\BibitemShut {NoStop}%
\bibitem [{\citenamefont {Meevasana}\ \emph {et~al.}(2008)\citenamefont
  {Meevasana}, \citenamefont {Baumberger}, \citenamefont {Tanaka},
  \citenamefont {Schmitt}, \citenamefont {Dunkel}, \citenamefont {Lu},
  \citenamefont {Mo}, \citenamefont {Eisaki},\ and\ \citenamefont
  {Shen}}]{Meevasana2008}%
  \BibitemOpen
  \bibfield  {author} {\bibinfo {author} {\bibfnamefont {W.}~\bibnamefont
  {Meevasana}}, \bibinfo {author} {\bibfnamefont {F.}~\bibnamefont
  {Baumberger}}, \bibinfo {author} {\bibfnamefont {K.}~\bibnamefont {Tanaka}},
  \bibinfo {author} {\bibfnamefont {F.}~\bibnamefont {Schmitt}}, \bibinfo
  {author} {\bibfnamefont {W.~R.}\ \bibnamefont {Dunkel}}, \bibinfo {author}
  {\bibfnamefont {D.~H.}\ \bibnamefont {Lu}}, \bibinfo {author} {\bibfnamefont
  {S.-K.}\ \bibnamefont {Mo}}, \bibinfo {author} {\bibfnamefont
  {H.}~\bibnamefont {Eisaki}}, \ and\ \bibinfo {author} {\bibfnamefont {Z.-X.}\
  \bibnamefont {Shen}},\ }\href
  {http://link.aps.org/doi/10.1103/PhysRevB.77.104506} {\bibfield  {journal}
  {\bibinfo  {journal} {Phys. Rev. B}\ }\textbf {\bibinfo {volume} {77}},\
  \bibinfo {pages} {104506} (\bibinfo {year} {2008})}\BibitemShut {NoStop}%
\bibitem [{\citenamefont {Levenberg}(1944)}]{levenberg}%
  \BibitemOpen
  \bibfield  {author} {\bibinfo {author} {\bibfnamefont {K.}~\bibnamefont
  {Levenberg}},\ }\href@noop {} {\bibfield  {journal} {\bibinfo  {journal}
  {Quart. Appl. Math.}\ }\textbf {\bibinfo {volume} {2}},\ \bibinfo {pages}
  {164} (\bibinfo {year} {1944})}\BibitemShut {NoStop}%
\bibitem [{\citenamefont {Marquardt}(1963)}]{marquardt}%
  \BibitemOpen
  \bibfield  {author} {\bibinfo {author} {\bibfnamefont {D.~W.}\ \bibnamefont
  {Marquardt}},\ }\href@noop {} {\bibfield  {journal} {\bibinfo  {journal} {J.
  Soc. Indust. Appl. Math.}\ }\textbf {\bibinfo {volume} {11}},\ \bibinfo
  {pages} {431} (\bibinfo {year} {1963})}\BibitemShut {NoStop}%
\bibitem [{\citenamefont {Hüfner}\ \emph {et~al.}(1999)\citenamefont
  {Hüfner}, \citenamefont {Claessen}, \citenamefont {Reinert}, \citenamefont
  {Straub}, \citenamefont {Strocov},\ and\ \citenamefont
  {Steiner}}]{HUFNER1999191}%
  \BibitemOpen
  \bibfield  {author} {\bibinfo {author} {\bibfnamefont {S.}~\bibnamefont
  {Hüfner}}, \bibinfo {author} {\bibfnamefont {R.}~\bibnamefont {Claessen}},
  \bibinfo {author} {\bibfnamefont {F.}~\bibnamefont {Reinert}}, \bibinfo
  {author} {\bibfnamefont {T.}~\bibnamefont {Straub}}, \bibinfo {author}
  {\bibfnamefont {V.}~\bibnamefont {Strocov}}, \ and\ \bibinfo {author}
  {\bibfnamefont {P.}~\bibnamefont {Steiner}},\ }\href {\doibase
  https://doi.org/10.1016/S0368-2048(99)00047-X} {\bibfield  {journal}
  {\bibinfo  {journal} {Journal of Electron Spectroscopy and Related
  Phenomena}\ }\textbf {\bibinfo {volume} {100}},\ \bibinfo {pages} {191 }
  (\bibinfo {year} {1999})}\BibitemShut {NoStop}%
\bibitem [{\citenamefont {Varma}\ \emph {et~al.}(1989)\citenamefont {Varma},
  \citenamefont {Littlewood}, \citenamefont {Schmitt-Rink}, \citenamefont
  {Abrahams},\ and\ \citenamefont {Ruckenstein}}]{Varma1989}%
  \BibitemOpen
  \bibfield  {author} {\bibinfo {author} {\bibfnamefont {C.~M.}\ \bibnamefont
  {Varma}}, \bibinfo {author} {\bibfnamefont {P.~B.}\ \bibnamefont
  {Littlewood}}, \bibinfo {author} {\bibfnamefont {S.}~\bibnamefont
  {Schmitt-Rink}}, \bibinfo {author} {\bibfnamefont {E.}~\bibnamefont
  {Abrahams}}, \ and\ \bibinfo {author} {\bibfnamefont {A.~E.}\ \bibnamefont
  {Ruckenstein}},\ }\href
  {https://link.aps.org/doi/10.1103/PhysRevLett.63.1996} {\bibfield  {journal}
  {\bibinfo  {journal} {PRL}\ }\textbf {\bibinfo {volume} {63}},\ \bibinfo
  {pages} {1996} (\bibinfo {year} {1989})}\BibitemShut {NoStop}%
\bibitem [{\citenamefont {Varma}\ \emph {et~al.}(1990)\citenamefont {Varma},
  \citenamefont {Littlewood}, \citenamefont {Schmitt-Rink}, \citenamefont
  {Abrahams},\ and\ \citenamefont {Ruckenstein}}]{varma1}%
  \BibitemOpen
  \bibfield  {author} {\bibinfo {author} {\bibfnamefont {C.~M.}\ \bibnamefont
  {Varma}}, \bibinfo {author} {\bibfnamefont {P.~B.}\ \bibnamefont
  {Littlewood}}, \bibinfo {author} {\bibfnamefont {S.}~\bibnamefont
  {Schmitt-Rink}}, \bibinfo {author} {\bibfnamefont {E.}~\bibnamefont
  {Abrahams}}, \ and\ \bibinfo {author} {\bibfnamefont {A.~E.}\ \bibnamefont
  {Ruckenstein}},\ }\href {\doibase 10.1103/PhysRevLett.64.497} {\bibfield
  {journal} {\bibinfo  {journal} {Phys. Rev. Lett.}\ }\textbf {\bibinfo
  {volume} {64}},\ \bibinfo {pages} {497} (\bibinfo {year} {1990})}\BibitemShut
  {NoStop}%
\bibitem [{\citenamefont {Zaanen}(2004)}]{planckian_1}%
  \BibitemOpen
  \bibfield  {author} {\bibinfo {author} {\bibfnamefont {J.}~\bibnamefont
  {Zaanen}},\ }\href {\doibase 10.1038/430512a} {\bibfield  {journal} {\bibinfo
   {journal} {Nature}\ }\textbf {\bibinfo {volume} {430}},\ \bibinfo {pages}
  {512} (\bibinfo {year} {2004})}\BibitemShut {NoStop}%
\bibitem [{\citenamefont {Zaanen}(2019)}]{planckian_2}%
  \BibitemOpen
  \bibfield  {author} {\bibinfo {author} {\bibfnamefont {J.}~\bibnamefont
  {Zaanen}},\ }\href {\doibase 10.21468/SciPostPhys.6.5.061} {\bibfield
  {journal} {\bibinfo  {journal} {SciPost Phys.}\ }\textbf {\bibinfo {volume}
  {6}},\ \bibinfo {pages} {61} (\bibinfo {year} {2019})}\BibitemShut {NoStop}%
\bibitem [{\citenamefont {Fink}\ \emph {et~al.}(2021)\citenamefont {Fink},
  \citenamefont {Rienks}, \citenamefont {Yao}, \citenamefont {Kurleto},
  \citenamefont {Bannies}, \citenamefont {Aswartham}, \citenamefont {Morozov},
  \citenamefont {Wurmehl}, \citenamefont {Wolf}, \citenamefont {Hardy},
  \citenamefont {Meingast}, \citenamefont {Jeevan}, \citenamefont {Maiwald},
  \citenamefont {Gegenwart}, \citenamefont {Felser},\ and\ \citenamefont
  {B\"uchner}}]{linkage}%
  \BibitemOpen
  \bibfield  {author} {\bibinfo {author} {\bibfnamefont {J.}~\bibnamefont
  {Fink}}, \bibinfo {author} {\bibfnamefont {E.~D.~L.}\ \bibnamefont {Rienks}},
  \bibinfo {author} {\bibfnamefont {M.}~\bibnamefont {Yao}}, \bibinfo {author}
  {\bibfnamefont {R.}~\bibnamefont {Kurleto}}, \bibinfo {author} {\bibfnamefont
  {J.}~\bibnamefont {Bannies}}, \bibinfo {author} {\bibfnamefont
  {S.}~\bibnamefont {Aswartham}}, \bibinfo {author} {\bibfnamefont
  {I.}~\bibnamefont {Morozov}}, \bibinfo {author} {\bibfnamefont
  {S.}~\bibnamefont {Wurmehl}}, \bibinfo {author} {\bibfnamefont
  {T.}~\bibnamefont {Wolf}}, \bibinfo {author} {\bibfnamefont {F.}~\bibnamefont
  {Hardy}}, \bibinfo {author} {\bibfnamefont {C.}~\bibnamefont {Meingast}},
  \bibinfo {author} {\bibfnamefont {H.~S.}\ \bibnamefont {Jeevan}}, \bibinfo
  {author} {\bibfnamefont {J.}~\bibnamefont {Maiwald}}, \bibinfo {author}
  {\bibfnamefont {P.}~\bibnamefont {Gegenwart}}, \bibinfo {author}
  {\bibfnamefont {C.}~\bibnamefont {Felser}}, \ and\ \bibinfo {author}
  {\bibfnamefont {B.}~\bibnamefont {B\"uchner}},\ }\href {\doibase
  10.1103/PhysRevB.103.155119} {\bibfield  {journal} {\bibinfo  {journal}
  {Phys. Rev. B}\ }\textbf {\bibinfo {volume} {103}},\ \bibinfo {pages}
  {155119} (\bibinfo {year} {2021})}\BibitemShut {NoStop}%
\bibitem [{\citenamefont {Kevan}(1986)}]{kevan}%
  \BibitemOpen
  \bibfield  {author} {\bibinfo {author} {\bibfnamefont {S.~D.}\ \bibnamefont
  {Kevan}},\ }\href {\doibase 10.1103/PhysRevB.33.4364} {\bibfield  {journal}
  {\bibinfo  {journal} {Phys. Rev. B}\ }\textbf {\bibinfo {volume} {33}},\
  \bibinfo {pages} {4364} (\bibinfo {year} {1986})}\BibitemShut {NoStop}%
\bibitem [{\citenamefont {Rinott}\ \emph {et~al.}(2017)\citenamefont {Rinott},
  \citenamefont {Chashka}, \citenamefont {Ribak}, \citenamefont {Rienks},
  \citenamefont {Taleb-Ibrahimi}, \citenamefont {Fevre}, \citenamefont
  {Bertran}, \citenamefont {Randeria},\ and\ \citenamefont {Kanigel}}]{rinott}%
  \BibitemOpen
  \bibfield  {author} {\bibinfo {author} {\bibfnamefont {S.}~\bibnamefont
  {Rinott}}, \bibinfo {author} {\bibfnamefont {K.}~\bibnamefont {Chashka}},
  \bibinfo {author} {\bibfnamefont {A.}~\bibnamefont {Ribak}}, \bibinfo
  {author} {\bibfnamefont {E.}~\bibnamefont {Rienks}}, \bibinfo {author}
  {\bibfnamefont {A.}~\bibnamefont {Taleb-Ibrahimi}}, \bibinfo {author}
  {\bibfnamefont {P.}~\bibnamefont {Fevre}}, \bibinfo {author} {\bibfnamefont
  {F.}~\bibnamefont {Bertran}}, \bibinfo {author} {\bibfnamefont
  {M.}~\bibnamefont {Randeria}}, \ and\ \bibinfo {author} {\bibfnamefont
  {A.}~\bibnamefont {Kanigel}},\ }\href {\doibase 10.1126/sciadv.1602372}
  {\bibfield  {journal} {\bibinfo  {journal} {Science Advances}\ }\textbf
  {\bibinfo {volume} {3}},\ \bibinfo {pages} {e1602372} (\bibinfo {year}
  {2017})}\BibitemShut {NoStop}%
\bibitem [{\citenamefont {Dynes}\ \emph {et~al.}(1978)\citenamefont {Dynes},
  \citenamefont {Narayanamurti},\ and\ \citenamefont {Garno}}]{dynes_gap}%
  \BibitemOpen
  \bibfield  {author} {\bibinfo {author} {\bibfnamefont {R.~C.}\ \bibnamefont
  {Dynes}}, \bibinfo {author} {\bibfnamefont {V.}~\bibnamefont
  {Narayanamurti}}, \ and\ \bibinfo {author} {\bibfnamefont {J.~P.}\
  \bibnamefont {Garno}},\ }\href {\doibase 10.1103/PhysRevLett.41.1509}
  {\bibfield  {journal} {\bibinfo  {journal} {Phys. Rev. Lett.}\ }\textbf
  {\bibinfo {volume} {41}},\ \bibinfo {pages} {1509} (\bibinfo {year}
  {1978})}\BibitemShut {NoStop}%
\bibitem [{\citenamefont {Bianchi}\ \emph {et~al.}(2010)\citenamefont
  {Bianchi}, \citenamefont {Rienks}, \citenamefont {Lizzit}, \citenamefont
  {Baraldi}, \citenamefont {Balog}, \citenamefont {Hornek\ae{}r},\ and\
  \citenamefont {Hofmann}}]{Bianchi2010}%
  \BibitemOpen
  \bibfield  {author} {\bibinfo {author} {\bibfnamefont {M.}~\bibnamefont
  {Bianchi}}, \bibinfo {author} {\bibfnamefont {E.~D.~L.}\ \bibnamefont
  {Rienks}}, \bibinfo {author} {\bibfnamefont {S.}~\bibnamefont {Lizzit}},
  \bibinfo {author} {\bibfnamefont {A.}~\bibnamefont {Baraldi}}, \bibinfo
  {author} {\bibfnamefont {R.}~\bibnamefont {Balog}}, \bibinfo {author}
  {\bibfnamefont {L.}~\bibnamefont {Hornek\ae{}r}}, \ and\ \bibinfo {author}
  {\bibfnamefont {P.}~\bibnamefont {Hofmann}},\ }\href
  {http://link.aps.org/doi/10.1103/PhysRevB.81.041403} {\bibfield  {journal}
  {\bibinfo  {journal} {Phys. Rev. B}\ }\textbf {\bibinfo {volume} {81}},\
  \bibinfo {pages} {041403} (\bibinfo {year} {2010})}\BibitemShut {NoStop}%
\bibitem [{\citenamefont {Shin}\ \emph {et~al.}(2020)\citenamefont {Shin},
  \citenamefont {Jung}, \citenamefont {Sohn}, \citenamefont {Ryu},
  \citenamefont {Huh},\ and\ \citenamefont {Kim}}]{shin_Rb}%
  \BibitemOpen
  \bibfield  {author} {\bibinfo {author} {\bibfnamefont {W.~J.}\ \bibnamefont
  {Shin}}, \bibinfo {author} {\bibfnamefont {S.~W.}\ \bibnamefont {Jung}},
  \bibinfo {author} {\bibfnamefont {Y.}~\bibnamefont {Sohn}}, \bibinfo {author}
  {\bibfnamefont {S.~H.}\ \bibnamefont {Ryu}}, \bibinfo {author} {\bibfnamefont
  {M.}~\bibnamefont {Huh}}, \ and\ \bibinfo {author} {\bibfnamefont {K.~S.}\
  \bibnamefont {Kim}},\ }\href {\doibase
  https://doi.org/10.1016/j.cap.2020.01.010} {\bibfield  {journal} {\bibinfo
  {journal} {Current Applied Physics}\ }\textbf {\bibinfo {volume} {20}},\
  \bibinfo {pages} {484} (\bibinfo {year} {2020})}\BibitemShut {NoStop}%
\bibitem [{\citenamefont {Engelsberg}\ and\ \citenamefont
  {Schrieffer}(1963)}]{fonony}%
  \BibitemOpen
  \bibfield  {author} {\bibinfo {author} {\bibfnamefont {S.}~\bibnamefont
  {Engelsberg}}\ and\ \bibinfo {author} {\bibfnamefont {J.~R.}\ \bibnamefont
  {Schrieffer}},\ }\href {\doibase 10.1103/PhysRev.131.993} {\bibfield
  {journal} {\bibinfo  {journal} {Phys. Rev.}\ }\textbf {\bibinfo {volume}
  {131}},\ \bibinfo {pages} {993} (\bibinfo {year} {1963})}\BibitemShut
  {NoStop}%
\bibitem [{\citenamefont {Fink}\ \emph {et~al.}(2007)\citenamefont {Fink},
  \citenamefont {Borisenko}, \citenamefont {Kordyuk}, \citenamefont {Koitzsch},
  \citenamefont {Geck}, \citenamefont {Zabolotnyy}, \citenamefont {Knupfer},
  \citenamefont {B{\"u}chner},\ and\ \citenamefont {Berger}}]{Fink2007}%
  \BibitemOpen
  \bibfield  {author} {\bibinfo {author} {\bibfnamefont {J.}~\bibnamefont
  {Fink}}, \bibinfo {author} {\bibfnamefont {S.}~\bibnamefont {Borisenko}},
  \bibinfo {author} {\bibfnamefont {A.}~\bibnamefont {Kordyuk}}, \bibinfo
  {author} {\bibfnamefont {A.}~\bibnamefont {Koitzsch}}, \bibinfo {author}
  {\bibfnamefont {J.}~\bibnamefont {Geck}}, \bibinfo {author} {\bibfnamefont
  {V.}~\bibnamefont {Zabolotnyy}}, \bibinfo {author} {\bibfnamefont
  {M.}~\bibnamefont {Knupfer}}, \bibinfo {author} {\bibfnamefont
  {B.}~\bibnamefont {B{\"u}chner}}, \ and\ \bibinfo {author} {\bibfnamefont
  {H.}~\bibnamefont {Berger}},\ }\enquote {\bibinfo {title} {Dressing of the
  charge carriers in high-t$_c$ superconductors},}\ in\ \href {\doibase
  10.1007/3-540-68133-7_11} {\emph {\bibinfo {booktitle} {Very High Resolution
  Photoelectron Spectroscopy}}},\ \bibinfo {editor} {edited by\ \bibinfo
  {editor} {\bibfnamefont {S.}~\bibnamefont {H{\"u}fner}}}\ (\bibinfo
  {publisher} {Springer Berlin Heidelberg},\ \bibinfo {address} {Berlin,
  Heidelberg},\ \bibinfo {year} {2007})\ pp.\ \bibinfo {pages}
  {295--325}\BibitemShut {NoStop}%
\bibitem [{\citenamefont {Hofmann}\ \emph {et~al.}(2009)\citenamefont
  {Hofmann}, \citenamefont {Sklyadneva}, \citenamefont {Rienks},\ and\
  \citenamefont {Chulkov}}]{Hofmann_2009}%
  \BibitemOpen
  \bibfield  {author} {\bibinfo {author} {\bibfnamefont {P.}~\bibnamefont
  {Hofmann}}, \bibinfo {author} {\bibfnamefont {I.~Y.}\ \bibnamefont
  {Sklyadneva}}, \bibinfo {author} {\bibfnamefont {E.~D.~L.}\ \bibnamefont
  {Rienks}}, \ and\ \bibinfo {author} {\bibfnamefont {E.~V.}\ \bibnamefont
  {Chulkov}},\ }\href {\doibase 10.1088/1367-2630/11/12/125005} {\bibfield
  {journal} {\bibinfo  {journal} {New Journal of Physics}\ }\textbf {\bibinfo
  {volume} {11}},\ \bibinfo {pages} {125005} (\bibinfo {year}
  {2009})}\BibitemShut {NoStop}%
\end{thebibliography}%

\end{document}